\documentclass[aip,jcp,preprint]{revtex4-1} 

\usepackage{amsmath}
\usepackage{amssymb}
\usepackage{algorithm}
\usepackage{algpseudocode}
\usepackage{amsfonts}
\usepackage{appendix}
\usepackage{amsthm}

\usepackage[dvips]{graphicx}
\usepackage{subfigure}
\usepackage{multirow}
\usepackage{fullpage}
\usepackage{graphicx,color}
\usepackage{soul}
\usepackage[normalem]{ulem}

\newcommand{\comment}[1]{}
\renewcommand{\sout}{\comment}

\begin{document}

\title{Topology of Classical Molecular Optimal Control Landscapes in Phase Space}
\author{Carlee Joe-Wong}
\altaffiliation{Present affiliation: Program in Applied and Computational Mathematics, Princeton University}
\affiliation{Department of Mathematics, Princeton University, Princeton, NJ, USA}
\author{Tak-San Ho}
\affiliation{Department of Chemistry, Princeton University, Princeton, NJ, USA}
\author{Ruixing Long}
\affiliation{Department of Chemistry, Princeton University, Princeton, NJ, USA}
\author{Herschel Rabitz}
\email{hrabitz@princeton.edu}
\affiliation{Department of Chemistry, Princeton University, Princeton, NJ, USA}
\author{Rebing Wu}
\affiliation{Department of Automation, Tsinghua University, Beijing, P.R. China}


\begin{abstract}
Optimal control of molecular dynamics is commonly expressed from a quantum mechanical perspective.  However, in most contexts the preponderance of molecular dynamics studies utilize classical mechanical models.  This paper treats laser-driven optimal control of molecular dynamics in a classical framework.  We consider the objective of steering a molecular system from an initial point in phase space to a target point, subject to the dynamic constraint of Hamilton's equations.  The classical control landscape corresponding to this objective is a functional of the control field, and the topology of the landscape is analyzed through its gradient and Hessian with respect to the control.  Under specific assumptions on the regularity of the control fields, the classical control landscape is found to be free of traps that could hinder reaching the objective.  The Hessian associated with an optimal control field is shown to have finite rank, indicating the presence of an inherent degree of robustness to control noise.  Extensive numerical simulations are performed to illustrate the theoretical principles on a) a model diatomic molecule, b) two coupled Morse oscillators, and c) a chaotic system with a coupled quartic oscillator, confirming the absence of traps in the classical control landscape.  We compare the classical formulation with the mathematically analogous \sout{quantum} state-to-state transition probability control landscape {\color{black} of $N$-level quantum systems. The absence of traps in both circumstances provides a broader basis to understand the growing number of successful control experiments with complex molecules, which can have dynamics that transcend the classical and quantum regimes.}
\end{abstract}

\maketitle

\section{Introduction} \label{sec:intro}

The control of molecular dynamics phenomena with tailored laser pulses is an active research field, and the common perspective is to view the dynamics in a quantum mechanical context \cite{brif2010control,rabitz2000whither,warren1993coherent}.  That situation is generally necessary when electronic degrees of freedom are involved.  However, molecular rotational and vibrational motion can often be reliably described classically, as frequently done for treating large polyatomic molecules.  Some prior works considered optimally controlled molecular dynamics using classical mechanics, and in some cases the results suggest that the classical and quantum pictures should have qualitatively similar behavior \cite{de2008quantum,gerber2000reflection,krempl1992optimal,miller1978classical,schwieters1991optimal,schwieters1993optimal,walker1977quantum}.  The conceptual and computational advantages of classically modeled control warrant further analysis of the foundations of the subject.  The present paper considers optimal classical control for steering a system from an initial point in phase space to a final target point.  This work builds on previous classical molecular optimal control studies \cite{schwieters1991optimal,schwieters1993optimal,batista2002coherent,botina1995new,botina1995optimal,demiralp1994optimal,efimov2003feedback,rice2009advances,krause1993classical,lissak1993optimal,nishiyama2004optimal,umeda2000quantum,taylor1982dynamical,yan1993optical} {\color{black} and extensive research into quantum control.\cite{brif2010control,rabitz2000whither,warren1993coherent,gerber2000reflection,ho2006effective,rabitz2006topology,rabitz2004quantum,hsieh2009topology,wu2008characterization,wu2010critical,pechen2010unified}}

An important feature of optimal control analysis is consideration of the underlying control landscape, defined as the physical objective as a functional of the control field.  At the \emph{critical points} of the landscape, the functional derivative of the objective with respect to the control is zero.  It is important to assess whether any critical points correspond to suboptimal traps on the landscape, which might prevent a search from reaching the absolute maximum value of the landscape.  It has recently been shown that under specified conditions the control landscape for maximizing the quantum state-to-state transition probability contains no traps,\cite{ho2006effective,rabitz2006topology,rabitz2004quantum} and the same conclusion holds for more general physical observables.~\cite{hsieh2009topology,wu2008characterization}  These latter analyses were stimulated by the observed ready achievement of effective control in broad classes of control simulations and experiments.  The landscape for optimizing the dynamical transformations in a classical system with harmonic potentials has been discussed on the symplectic group, and was shown to be devoid of traps.\cite{wu2010critical} A recent paper \cite{pechen2010unified} examined the optimal control landscape of open classical and quantum systems, showing that traps do not exist under well-defined conditions.  The latter work relies on the open nature of the systems, and the present paper considers the dynamics of closed systems.

In the case of quantum mechanics, prior analyses considered the control landscape for the probability of making a transition from one pure quantum state to another ($|i\rangle\rightarrow|f\rangle$) in a system of $N$ energy levels. Here we treat the classical mechanical analogue of making a ``pure phase-space state transition'' ${\bf z}\rightarrow {\bf z'}$ for a system of $n$ particles (i.e., ${\bf z}$ is a point in the coordinate-momentum phase space).  Drawing an analogy between a quantum system of $N$ states and a classical system of $n$ particles is unusual, but we show that they share strikingly similar landscape topologies.  We thus extend previous works that formulate classical and quantum mechanics as analogues, either in terms of a classical Hamiltonian system \cite{mendes2003quantum, heslot1985quantum} or Poisson bracket operators.\cite{shirokov1979quantum}  The formulations discussed in these papers do not consider the control landscape topology. The goal of {\color{black} this} work is to (i) analyze the structure of the classical phase space control landscape and (ii) perform numerical simulations supporting the theoretical conclusions about the landscape's properties. {\color{black} In doing so, we aim to provide a foundation to understand the success of experimental control studies on complex molecules which likely exhibit classical behavior.}

In some contexts, the potential for chaotic behavior of classical molecular systems may present difficulties for optimal control.  However, some studies have shown that under control, even systems that are nominally chaotic may exhibit  regular, non-chaotic behavior.\cite{botina1995optimal, schwieters1991optimal, taylor1982dynamical} Managing chaotic behavior is relevant for the stability of optimally controlled trajectories.  In this work, we consider the use of open loop control to find such an optimal trajectory, with its stability assessed \emph{post-facto}.  Other works \cite{cai2002converting,lai1993controlling, ott1990controlling,wu1998controlling} have considered using active feedback control to stabilize chaotic systems about an optimal trajectory. While there is no theoretical barrier to finding an optimally controlled trajectory in the chaotic regime, it can be numerically difficult due to chaotic systems' high sensitivity to initial conditions. An alternative procedure could redefine the cost function, for example, as an average over a window centered around a target time or an ensemble average over initial conditions.  In the present work, we numerically illustrate a nominally chaotic coupled quartic oscillator system over several periods of motion, which is consistent with many analogous quantum control problems{\color{black} . The simulations reflect the underlying chaotic dynamics while operating in a  numerically tractable regime for assessing the landscape principles. Further exploration of  chaotic controlled dynamics deserves special attention, which is beyond the scope of the present work.}

{\color{black} In Section \ref{sec:eqns} of the paper, we introduce} the classical system and the target functional $J$.  Section \ref{sec:criticalpoints} considers the control landscape critical points where the first-order functional derivative of $J$ with respect to the control field is zero.  Section \ref{sec:hessian} then analyzes the Hessian of the control landscape at the critical points, while Section \ref{sec:stability} discusses the robustness and multiplicity of optimal solutions by examining the spectrum of the Hessian.  The findings in Sections \ref{sec:criticalpoints} and \ref{sec:hessian} are compared in Section \ref{sec:quantum} with analogous quantum mechanical results for maximizing pure state transition probabilities. Section \ref{sec:simulations} presents numerical simulations for a {\color{black} range of simple model systems} to illustrate the results in Section \ref{sec:eqns}. We discuss simulation results for a one-dimensional Morse oscillator and then consider two coupled Morse oscillators {\color{black} before turning to a coupled quartic oscillator that exhibits chaotic behavior.} Concluding remarks {\color{black} on the work's significance and future extensions} are given in Section \ref{sec:conclusion}.

\section{Optimal phase space control within classical mechanics}\label{sec:eqns}

Consider a classically described molecule of $n$ atoms, driven by a linearly polarized electric control field $\epsilon(t)$.  The Hamiltonian $H({\bf p},{\bf q},t)$ of the polyatomic molecule can be written as
\begin{equation}
H({\bf p},{\bf q},t) = \frac{1}{2}{\bf p}^T{\bf M}^{-1}{\bf p} + V({\bf q}) - D({\bf q})\epsilon(t),
\label{eq:hamiltonianfree}
\end{equation}
where ${\bf p}={\bf p}(t)$ and ${\bf q}={\bf q}(t)$ are, respectively, $3n$-dimensional momentum and position vectors describing the
state of the $n$ atoms in a three-dimensional Cartesian coordinate frame and
\begin{equation}
{\bf M} = \begin{bmatrix} {\bf m}_1  & 0 & \ldots & 0 \\ 0 & {\bf m}_2 & \ldots & 0 \\ \vdots & \vdots & \ddots & \vdots \\ 0 & 0 & \ldots & {\bf m}_n \end{bmatrix},
\end{equation}
is a $3n\times 3n$ block diagonal matrix composed of $3\times 3$ sub-blocks
\begin{equation}
{\bf m}_i = \begin{bmatrix} m_i & 0 & 0 \\ 0 & m_i & 0 \\ 0 & 0 & m_i \end{bmatrix};\;i = 1,2,\ldots, n,
\end{equation}
with $m_i$ being the mass of the $i$-th atom.  The functions $V({\bf q})$ and $D({\bf q})$ are, respectively,
the potential energy between the atoms and the projection of the dipole moment along the control field $\epsilon(t)$.  The analysis below assumes that $V({\bf q})$ and $D({\bf q})$ are twice differentiable functions of the coordinates ${\bf q}$
and that $\epsilon(t)$ is square-integrable on the time interval $[0,T]$.  For an arbitrary polyatomic molecule, the $3n$ Cartesian coordinates ${\bf q}$ may be further written in terms of three center-of-mass coordinates, three overall rotational angular coordinates, and $3n-6$ internal coordinates.  For simplicity, we do not make the latter transformations, and for ease of notation, we set ${\cal N} = 3n$. {\color{black} Reduced cases of the treatment above apply for motion in two or one dimension(s) when $\mathcal{N} = 2n$ and $n$, respectively.}

The column vector ${\bf z}(t) = \begin{bmatrix} {\bf q}^T(t) & {\bf p}^T(t) \end{bmatrix}^T$  
specifies the state of the molecule at time $t$ in the corresponding $2{\cal N}$-dimensional phase space.  The dynamical evolution of the molecule is then governed by Hamilton's equations:
\begin{equation}
\dot{{\bf z}} = \begin{bmatrix} \dot{{\bf q}} \\ \dot{{\bf p}} \end{bmatrix} = \begin{bmatrix} \left(\frac{\partial H({\bf p},{\bf q},t)}{\partial {\bf p}}\right)^T \\ -\left(\frac{\partial H({\bf p},{\bf q},t)}{\partial {\bf q}}\right)^T\end{bmatrix}= \begin{bmatrix} 0 & I \\ -I & 0 \end{bmatrix}
 \begin{bmatrix} \frac{\partial H({\bf p},{\bf q},t)}{\partial {\bf q}} & \frac{\partial H({\bf p},{\bf q},t)}{\partial {\bf p}}\end{bmatrix}^T,
\label{eq:sysdym}
\end{equation}
where the dot denotes the time derivative and $I$ is the $\mathcal{N}\times\mathcal{N}$ identity matrix.  Equation (\ref{eq:sysdym}) possesses a unique solution \cite{sontag1998mathematical}, given an
initial state ${\bf z_0} = \begin{bmatrix} {\bf q_0}^T & {\bf p_0}^T \end{bmatrix}^T$.
We assume that Eq. (\ref{eq:sysdym}) has a well-defined solution for $t\in[0, T]$.  We further make the physically reasonable assumption that the potential and dipole functions are bounded, at least over the accessible dynamics on the interval $[0,T]$.  {\color{black} Unbounded potentials such as the Coulomb potential} \sout{would thus }can also {\color{black} satisfy our assumptions (e.g., the Coulomb potential has been used to classically model hydrogen atom ionization\cite{abrines1966classical}).}  We assume that the finite time interval implies that all states ${\bf z}(t)$ driven by admissible control fields are uniformly bounded over $t\in[0,T]$ (i.e., the system exhibits no finite-time blowups).  Thus, an unbounded potential would satisfy our assumption, as long as the points at which it is unbounded do not lie within the sampled phase space. The objective is to seek an optimal control field to steer the molecule from its initial state ${\bf z}_0$
to a final state ${\bf z}(T)$, in order to maximize a specified function $O({\bf z}(T))$.  This function is constructed to increase as ${\bf z}(T)$ comes close to the desired target state ${\bf z_{\rm tar}}$ in phase space.  Thus, we seek to assess the maximization of the objective functional $J$
\begin{equation} \label{eq:cost}
\max_{\epsilon(\cdot)} J\equiv O({\bf z}(T))
\end{equation}
with respect to the control field  $\epsilon(t)$, $t\in[0,T]$.  The {\it control landscape} $J\equiv J[\epsilon(t)]$ is a functional of the control field.  We assume that all final states ${\bf z}(T)$, including those that maximize $O({\bf z}(T))$, lie in the interior of the \emph{reachable set} from ${\bf z_0}$ at time $T$.  The latter set consists of all final states at time $T$ to which the initial state ${\bf z_0}$ may be steered for some admissible control field $\epsilon(t)$ [27-30].\nocite{bonnard2003singular,geometric,sontag1998mathematical}{\tiny{\color{white}\footnote{Sufficient conditions are known under which the reachable set has a non-empty interior.  However, it is difficult to derive general necessary and sufficient conditions \cite{geometric}.}}}Moreover, for simplicity, $O({\bf z}(T))$ is assumed to be twice continuously differentiable with respect to ${\bf z}(T)$ and bounded from above (see Section \ref{sec:hessian}).  Other terms may be included in the objective functional $J$ (e.g., $J\equiv O({\bf z}(T))+\alpha\int_0^T{\epsilon(t)^2\;dt}$, $\alpha>0$), to bias the class of controls.  As in the quantum mechanical analysis,\cite{rabitz2004quantum,rabitz2006topology} here we focus on the fundamental case where $J[\epsilon(t)] = O({\bf z}(T))$ to assess the landscape topology without additional objectives or constraints on the field. {\color{black} Choosing \sout{The fact that }}$O(z(T))$ {\color{black} so that it} rises as ${\bf z}(T)\rightarrow {\bf z}_{\rm tar}$ does not in itself assure that $J$ is trap-free with respect to $\epsilon(t)$.  Analyzing the behavior of $J$ with respect to $\epsilon(t)$ is an important goal of this paper.

\subsection{Critical points of the control landscape}\label{sec:criticalpoints}

The critical points of the control landscape $J[\epsilon(t)]$ correspond to the zero gradient condition
\begin{equation}
\frac{\delta J}{\delta \epsilon(t)} = \frac{\partial J}{\partial {\bf z}(T)}\frac{\delta {\bf z}(T)}{\delta \epsilon(t)}=0,\;\forall\,t \in[0,T].
\label{eq:chain}
\end{equation}
The control landscape critical points may be divided into two types: kinematic critical points, at which $\partial J/\partial {\bf z}(T) = 0$, and non-kinematic critical points, at which $\partial J/\partial {\bf z}(T) \neq 0$.\cite{wu2012singularities}  Kinematic critical points may be further considered as regular, at which the 
{\color{black} functional mapping $\delta\epsilon(\cdot)\rightarrow\delta{\bf z}(T)$} is of full rank (often referred to as being surjective), or singular, at which 
is not of full rank.\cite{sontag1998mathematical}  A regular control $\epsilon(t)$ is {\color{black} thus} defined as having the property that an appropriate perturbation $\epsilon(t)\rightarrow\epsilon(t) + \delta\epsilon(t)$ can move the final state ${\bf z}(T)\rightarrow{\bf z}(T) + \delta{\bf z}(T)$ in any specified direction.

We assume that {\color{black} the functional mapping $\delta\epsilon(\cdot)\rightarrow \delta{\bf z}(T)$} 
is of full rank (i.e., {\color{black} the vector-valued function of time} $\delta{\bf z}(T)/\delta\epsilon(t)$ is of full rank $2{\cal N}\; (=6n)$) for \sout{any}{\color{black} the} field $\epsilon(t)$ {\color{black} at the critical point of the control landscape under consideration}.  Using Eq. (\ref{eq:chain}) it then follows that $\partial J/\partial {\bf z}(T) = 0$ at a critical point.  Numerical simulations (see Section \ref{sec:simulations}) suggest that singular controls are rare, which is consistent with the assumption of typical fields being regular and $\delta{\bf z}(T)/\delta\epsilon(t)$ being surjective.  We will also assess the likely existence of regular controls through an explicit expression derived below for $\delta{\bf z}(t)/\delta\epsilon(t)$.  A bound on $\delta{\bf z}(t)/\delta\epsilon(t)$ will also be identified, which will be used in Section \ref{sec:stability} to explore the robustness of the control solutions to noise at the critical points.

Consider a perturbation of the control $\epsilon(t)\rightarrow\epsilon(t) + \delta\epsilon(t)$ in Eq. (\ref{eq:sysdym}) \cite{bonnard2003singular} with the corresponding response in ${\bf z}(t)\rightarrow {\bf z}(t) + \delta{\bf z}(t)$ leading to the first order variation
\begin{eqnarray}
{\bf \delta\dot{z}}(t)
&=& A(t){\bf \delta z}(t) + B(t)\delta \epsilon(t),
\label{eq:changex}
\end{eqnarray}
where $\delta{{\bf z}}^T(t) = \begin{bmatrix} {\delta{\bf q}}^T(t) & {\delta{\bf p}}^T(t) \end{bmatrix}^T$,
\begin{align}
A(t) &= \begin{bmatrix} \emptyset_{{\cal N}} & {\bf M}^{-1} \\ -\frac{\partial^2 V({\bf q})}{\partial {\bf q}^2} + \frac{\partial^2 D({\bf q})}{\partial {\bf q}^2}\epsilon(t) & \emptyset_{{\cal N}} \end{bmatrix}, \label{eq:A}
\end{align}
and
\begin{align}
B(t) &= \begin{bmatrix} 0\\ \vdots\\0 \\ -\left(\frac{\partial D({\bf q})}{\partial {\bf q}}\right)^T \end{bmatrix} \label{eq:B},
\end{align}
Here $\emptyset_{{\cal N}}$ is the ${\cal N}\times {\cal N}$ square zero matrix in $A(t)$, and the first ${\cal N}$ entries of $B(t)$ are zero.  The notation $\partial^2 X/\partial{\bf q}^2$ in $A(t)$ refers to the following ${\cal N}\times {\cal N}$ second derivative matrix of the function $X(\bf{q})$ (i.e., for $V({\bf q})$ or $D({\bf q})$):
\begin{equation}
\frac{\partial^2 X}{\partial{\bf q}^2} =\frac{\partial}{\partial{\bf q}}\left(\frac{\partial X}{\partial{\bf q}}\right)
= \begin{bmatrix} \frac{\partial^2 X}{\partial q_1^2} & \frac{\partial^2 X}{\partial q_1\partial q_2} & \ldots & \frac{\partial^2 X}{\partial q_1\partial q_{{\cal N}}} \\ \frac{\partial^2 X}{\partial q_2\partial q_1} & \frac{\partial^2 X_2}{\partial q_2^2} & \ldots & \frac{\partial^2 X}{\partial q_1\partial q_{{\cal N}}} \\ \vdots & \vdots & \ddots & \vdots \\ \frac{\partial^2 X}{\partial q_{{\cal N}}\partial q_1} & \frac{\partial^2 X}{\partial q_{{\cal N}}\partial q_2} & \ldots & \frac{\partial^2 X}{\partial q_{{\cal N}}^2} \end{bmatrix}.
\end{equation}
Similarly, $\partial D({\bf q})/\partial {\bf q}$ is the ${\cal N}$-dimensional gradient row vector of $D({\bf q})$ with respect to the ${\cal N}$ components of ${\bf q}$.
Integrating Eq. (\ref{eq:changex}) yields
\begin{equation}\label{eq:deltaz}
{\bf\delta z}(T) = M(T)\int_0^T{M^{-1}(t)B(t)\delta\epsilon(t)\;dt},
\end{equation}
where $M(t)$ satisfies the equation~\cite{brockett1970finite} 
\begin{equation}
\dot{M}(t) = A(t)M(t),\quad M(0) = I.
\label{eq:Mdym}
\end{equation}
The elements of $M$ have the meaning $M_{ij}(t) = \partial z_i(t)/\partial z_j(0)$, representing the sensitivity of the $i$th component of the state at time $t$ to the $j$th component of the state at the initial time.  The matrix $M(t)$ is also symplectic \cite{dragt2005symplectic} and thus invertible.  From Eq. (\ref{eq:deltaz}) we have
\begin{equation}
\frac{\delta {\bf z}(T)}{\delta\epsilon(t)} = M(T)M^{-1}(t)B(t).
\label{eq:deltaze}
\end{equation}  

Based on Eq. (\ref{eq:deltaze}), the regularity criterion for a field $\epsilon(t)$ or equivalently the surjectivity of $\delta{\bf z}(T)/\delta\epsilon(t)$ corresponds to the condition that the vectors $M^{-1}(t)B(t)$ span $\mathbb{R}^{2{\cal N}}$, since $M(T)$ is an invertible matrix.  In the Appendix, we explicitly demonstrate regularity for critical point controls of a one-dimensional linear forced harmonic oscillator, for which the vector $M^{-1}(t)B(t)$ is of full rank (i.e., of rank $2$) even though the corresponding matrix $A(t)$ in Eq. (\ref{eq:Mdym}) is independent of time $t$.  For a general $n$-particle Hamiltonian system, the matrix $A(t)$ in Eq. (\ref{eq:A}) will likely be a complicated function of time (via the state variables ${\bf q}(t)$ and the field $\epsilon(t)$).  It appears difficult to prove surjectivity under all conditions, but for arbitrary Hamiltonians and fields the mere complexity of the dynamics suggest that the vector components of $M^{-1}(t)B(t)$, as a function of time $0\leq t\leq T$, are likely of full rank $2{\cal N}$. 
We adopt this assumption here and investigate its consequences, which {\color{black} will} be tested in numerical simulations {\color{black} in Section \ref{sec:simulations}}.

In Section \ref{sec:hessian}, we use the second-order functional derivative of $J$ with respect to the control to investigate the optimality of regular critical points and the associated robustness to field errors $\epsilon(t)\rightarrow\epsilon(t) + \delta\epsilon(t)$.  These investigations enable a \emph{post-facto} evaluation of the stability of the evolved system trajectory.  The latter analysis utilizes the time-averaged norm of the functional derivative $\delta{\bf z}(T)/\delta\epsilon(t)$:
\begin{equation}\label{eq:bounddzde}
\left\|\frac{\delta {\bf z}(T)}{\delta\epsilon(\cdot)}\right\|_T^2
\equiv \frac{1}{T}\int_0^T \left(\frac{\delta {\bf z}(T)}{\delta\epsilon(t)}\right)^T\frac{\delta {\bf z}(T)}{\delta\epsilon(t)}\;dt
\leq C^2\left\|B(\cdot)\right\|_T^2,
\end{equation}
where 
\begin{equation}
\left\|B(\cdot)\right\|_T^2\equiv\frac{1}{T}\int_0^T\left\|B(t)\right\|^2\;dt=\frac{1}{T}\int_0^T\frac{\partial D({\bf q})}{\partial{\bf q}}\left[\frac{\partial D({\bf q})}{\partial{\bf q}}\right]^T\;dt,
\label{eq:defB}
\end{equation}
which is expected to be finite for typical molecular dipole moments and bounded dynamics ${\bf q}(t)$, $0\leq t\leq T$.  Here $\|\cdot\|$ is the Euclidean norm of a vector and $\|\cdot\|_{\rm F}$ is the Frobenius norm of a square matrix.  Moreover, we have assumed that $\left\|M(T)M^{-1}(t)\right\|_{\rm F}$ is uniformly bounded (i.e., $\left\|M(T)M^{-1}(t)\right\|_{\rm F}\leq C$ for all $t\in[0,T]$, $C$ being a finite positive number).  The entries of $M^{-1}(t)$ may be bounded by exponentiating a uniform bound on the entries of $A(t)$ using Eq. (\ref{eq:Mdym}).

From Eq. (\ref{eq:A}), the existence of such a bound on the entries of $A$ corresponds to a uniform bound on the second derivatives of the potential and dipole moment functions.  Since we assume that the state trajectories are bounded and that $V({\bf q})$ and $D({\bf q})$ are twice continuously differentiable, an upper bound on the entries of $M^{-1}(t)$ exists.
Furthermore, we can then show that the time average of the gradient $\delta J/\delta\epsilon(t)$ is bounded as follows: 
\begin{equation}
\left|\frac{\delta J}{\delta\epsilon(\cdot)}\right|_T \equiv \frac{1}{T}\int_0^T \left|\frac{\partial J}{\partial {\bf z}(T)} \frac{\delta {\bf z}(T)}{\delta\epsilon(t)}\right|\;dt 
\leq \frac{C}{\sqrt{T}}\left\|\frac{\partial J}{\partial {\bf z}(T)}\right\| \times\left\|B(\cdot)\right\|_T.\label{eq:bound2}
\end{equation}
The result in Eq. (\ref{eq:bound2}) shows that extraordinarily steep regions should not exist on the landscape, thereby favoring the stability of algorithms seeking to optimize $J[\epsilon(t)]$.  Given these critical point conditions and gradient bounds, below we consider the optimality of regular critical point controls, using the second-order functional derivative of $J$ with respect to the control field.

\subsection{Nature of the control landscape critical points: Hessian analysis} \label{sec:hessian}

This section considers the second-order functional derivative ${\cal H}(t,t')\equiv\delta^2 J/\delta\epsilon(t)\delta\epsilon(t')$ (i.e., the Hessian of the objective with respect to the control) evaluated at the critical points.  The properties of $\mathcal{H}(t,t')$ determine the conditions under which the landscape may contain traps (i.e., controls producing local, suboptimal maxima).  Using the relation $\delta J/\delta\epsilon(t) = \left[\partial J/\partial {\bf z}(T)\right]\left[\delta {\bf z}(T)/\delta\epsilon(t)\right]$ and differentiating again, the Hessian ${\cal H}(t,t')$ can be written as
\begin{equation}
{\cal H}(t,t')= \frac{\partial J}{\partial {\bf z}(T)}\frac{\delta^2 {\bf z}(T)}{\delta\epsilon(t)\delta\epsilon(t')} + \left(\frac{\delta {\bf z}(T)}{\delta\epsilon(t')}\right)^T\frac{\partial^2 J}{\partial {\bf z}(T)^2}\frac{\delta {\bf z}(T)}{\delta\epsilon(t)},
\label{eq:hessian1}
\end{equation}
where \begin{equation*}\frac{\partial^2 J}{\partial {\bf z}(T)^2}\equiv\frac{\partial}{\partial{\bf z}(T)}\left(\frac{\partial J}{\partial {\bf z}(T)}\right).\end{equation*}
At kinematic critical points, $\partial J/\partial {\bf z}(T) = 0$ (see Section \ref{sec:criticalpoints}) and the Hessian becomes
\begin{equation} 
{\cal H}(t,t')= \frac{\delta^2 J}{\delta\epsilon(t)\delta\epsilon(t')} = \left(\frac{\delta {\bf z}(T)}{\delta\epsilon(t')}\right)^T\frac{\partial^2 J}{\partial {\bf z}(T)^2}\frac{\delta {\bf z}(T)}{\delta\epsilon(t)},
\label{eq:partial2}
\end{equation}
which is symmetric and separable, thus at most of finite rank $2{\cal N}$, dictated by $\partial^2 J/\partial{\bf z}(T)^2$.  Here we use the assumption that $J$ is twice continuously differentiable with respect to ${\bf z}(T)$ to conclude that $\partial^2 J/\partial {\bf z}(T)^2$ is a symmetric matrix.
 
Establishing the positive, negative or indefinite nature of the Hessian is essential for assessing whether traps exist on the landscape.  We can test for the character of the Hessian by diagonalizing the matrix $\partial^2 J/\partial {\bf z}(T)^2$; since this matrix is real and symmetric, there exists an orthogonal matrix ${\bf P}$ such that
\begin{equation}
\frac{\partial^2 J}{\partial {\bf z}(T)^2} = {\bf P^T\Sigma P},
\label{eq:jdiag}
\end{equation}
where ${\bf \Sigma}$ is a diagonal matrix, whose diagonal entries $\sigma_i$, $i = 1,2,\ldots, 2\mathcal{N}$ are the eigenvalues of $\partial^2 J/\partial {\bf z}(T)^2$ \cite{hoffman1971linear}.  Then Eq. (\ref{eq:partial2}) becomes
\begin{equation}
\mathcal{H}(t,t') = \left({\bf P}\frac{\delta {\bf z}(T)}{\delta\epsilon(t')}\right)^T{\bf \Sigma P}\frac{\delta {\bf z}(T)}{\delta\epsilon(t)},
\label{eq:hessianfinal}
\end{equation}
which may be re-written as
\begin{equation}
\mathcal{H}(t,t') = \displaystyle\sum_{l=1}^{2\mathcal{N}}{\sigma_l\phi_l(t)\phi_l(t')},
\label{eq:hessianexpansion}
\end{equation}
where each $\phi_l(t)$ is the $l$th component of the vector ${\bf P}\left(\delta{\bf z}(T)/\delta\epsilon(t)\right)$.  Thus, the rank of the Hessian $\mathcal{H}(t,t')$ equals the number of nonzero eigenvalues $\sigma_i$ of $\partial^2 J/\partial {\bf z}(T)^2$.  Moreover, {\color{black} it can be shown \cite{ho2009landscape} that} the signature of the Hessian (the difference between the number of positive and number of negative eigenvalues) is the signature of ${\bf \Sigma}$, or equivalently $\partial^2 J/\partial {\bf z}(T)^2$, from Eq. (\ref{eq:jdiag}).

Since $J\equiv O({\bf z}(T))$, we see that a regular control field $\epsilon(t)$ corresponds to a local or global maximum if it produces a final state ${\bf z}(T)$ such that $\partial^2 J/\partial {\bf z}(T)^2$ is negative definite.  Importantly, we are free to choose the observable $J = O({\bf z}(T))$ on physical grounds so that it contains only one maximum which is global, and in this case the control landscape contains no traps for regular control fields.  Moreover, the landscape will contain saddle points at regular control fields if and only if $O({\bf z}(T))$ has saddle points.  Thus, traps on the landscape under the assumptions of the analysis should only arise if they were artificially introduced into $O({\bf z}(T))$.

Establishing a bound on the Hessian is important (see Section \ref{sec:stability}), as it reflects the degree of robustness at the optimal value of $J$ to variations in the field $\epsilon(t)\rightarrow\epsilon(t) + \delta\epsilon(t)$ due to noise.  We can obtain a bound on the trace of the Hessian ${\cal H}(t,t')$ using Eq. (\ref{eq:hessianfinal}):
\begin{equation}
\left|{\rm Tr}{\,\cal H}\right| \equiv \left|\int_0^T {\cal H}(t,t)\;dt\right|
\leq C^2T \left\| B(\cdot)\right\|_T^2\times\sum_{l=1}^{2{\cal N}}\sigma_l. \label{eq:tracebound}
\end{equation}
For example, if $O({\bf z}(T))$ is the quadratic function $[{\bf z}(T)-{\bf z_{\rm tar}}]^T{\bf W}[{\bf z}(T)-{\bf z_{\rm tar}}]$, where ${\bf W}$ is a symmetric, negative-definite $2{\cal N}\times 2{\cal N}$ matrix, then $O({\bf z}(T))$ has only one critical value (the global maximum) $O({\bf z}(T))=0$, corresponding to ${\bf z}(T)-{\bf z_{\rm tar}} = 0$.  In this special case, it is readily seen that the kinematic Hessian at the global maximum is the $2{\cal N}\times 2{\cal N}$ matrix
$\partial^2 J/\partial {\bf z}(T)^2 =2{\bf W}$, which is negative
definite and of full rank $2{\cal N}$ by assumption.  Therefore ${\rm Tr}\,\mathcal{H}$ will be negative, as ${\rm Tr}\left(\partial^2 J/\partial {\bf z}(T)^2\right)=2{\rm Tr}\,{\bf W}< 0$.

\subsection{Robustness analysis and level sets at the control landscape global maximum} \label{sec:stability}

The existence of only a finite number of nonzero eigenvalues for ${\cal H}(t,t')$ at controls producing global maxima implies that the Hessian has infinitely many zero eigenvalues (i.e., ${\cal H}(t,t')$ has an infinite-dimensional nullspace).  This property has important consequences for the robustness to noise and the multiplicity of optimal controls.  Qualitatively, a control solution $\epsilon(t)$ is robust if a perturbation $\delta\epsilon(t)$ does not significantly change the value of $J$.  The fact that the Hessian possesses infinitely many zero eigenvalues indicates that the landscape is inherently flat at the global maximum value with the existence of a submanifold of optimal control solutions around a given optimal field.  

The robustness of an optimal control solution follows from the previous bounds on the trace of the Hessian in Eq. (\ref{eq:tracebound}).  Consider the response $\delta J$ due to a perturbation $\delta\epsilon(t)$ up to second-order
\begin{equation}\label{eq:deltaJ}
\delta J =\int_0^T\frac{\delta J}{\delta \epsilon(t)}\delta\epsilon(t)\;dt
+\frac{1}{2} \int_0^T{\int_0^T{\frac{\delta^2 J}{\delta\epsilon(t)\delta\epsilon(t')}\delta\epsilon(t)\delta\epsilon(t')\;dt'}\;dt}.
\end{equation}
Using Eq. (\ref{eq:hessianexpansion}) at the control maximum (i.e.,  $\delta J/\delta\epsilon(t)=0$),  Eq. (\ref{eq:deltaJ}) implies that
\begin{equation}\label{eq:boundJ}
\left|\delta J\right| =\frac{1}{2}\sum_{l=1}^{2{\cal N}} \left|\sigma_l\right| \left(\int_0^T\int_0^T\phi_l(t)\phi_l(t')\delta\epsilon(t)\delta\epsilon(t')\;dt\;dt'\right)^2
\leq\frac{T^2}{2} \|\delta\epsilon\|_T^2\times\sum_{l=1}^{2{\cal N}} \left|\sigma_l\right| \|\phi_l\|_T^2, 
\end{equation}
where we used $\sigma_l\leq 0;\ l = 1,2,\ldots,2{\cal N}$ for a control corresponding to the landscape's maximum value.  The norm $\|\delta\epsilon\|_T$ is defined as in Eq. (\ref{eq:defB}):
\begin{equation}
\|\delta\epsilon\|_T^2\equiv\frac{1}{T}\int_0^T\left[\delta\epsilon(t)\right]^2\;dt.
\end{equation} 
Since the trace of the Hessian ${\rm Tr}{\,\cal H}$ satisfies
\begin{equation}
{\rm Tr}{\,\cal H}=\int_0^T{\cal H}(t,t)\;dt=\sum_{l=1}^{2{\cal N}}\sigma_l\int_0^T\phi_l^2(t)\;dt=T\sum_{l=1}^{2{\cal N}}\sigma_l\|\phi_l\|_T^2,
\end{equation}
Eq. (\ref{eq:boundJ}) can be rewritten as
\begin{equation}
\left|\delta J\right| \leq\frac{T}{2} \|\delta\epsilon\|_T^2\times\left|{\rm Tr}{\,\cal H}\right|\leq \frac{1}{2}C^2T^2 \|\delta\epsilon\|_T^2\times \left\|B(\cdot)\right\|_T^2\times \left|{\rm Tr}\left(\frac{\partial^2 J}{\partial {\bf z}(T)^2}\right)\right|,
\end{equation}
where we used the fact that ${\rm Tr}\left(\partial^2 J/\partial {\bf z}(T)^2\right)< 0$.  This result shows that a near-optimal control field (i.e., $\epsilon(t) + \delta\epsilon(t)$) will produce a near-optimal objective functional value $J$.  The bound on the resultant variation $\delta J$ scales with the squared-norm of the control field and the magnitude of the trace of the second derivative of $J$ with respect to the final state.  This result along with the infinite-dimensional nullspace of $\mathcal{H}(t,t')$ implies the existence of inherent robustness to control noise at the landscape maximum value.

The analysis above also suggests that a level set of optimal controls exists near any given optimal solution $\epsilon(t)$, since the objective functional $J$ remains maximized when the control $\epsilon$ is systematically altered by a variation that lies in the infinite-dimensional nullspace of ${\cal H}(t,t')$.  To move the control in this nullspace, we adapt the procedure exploited for the analogous behavior of quantum systems.\cite{rothman2006exploring}  The control field is now denoted as $\epsilon(s,t)$, where the variable $s\geq 0$ labels a point in the evolution of the control field within the nullspace of ${\cal H}(t,t')$.  Since the gradient satisfies $\delta J/\delta \epsilon(s,t) = 0$ at the optimum, the first derivative $\partial J/\partial s$ is automatically zero,
\begin{equation*}
\frac{\partial J}{\partial s} = \int_0^T{\frac{\delta J}{\delta\epsilon(t)}\frac{\partial\epsilon(s,t)}{\partial s}\;dt} \equiv 0,
\end{equation*}
and we wish to choose $\partial\epsilon(s,t)/\partial s$ so that the second derivative satisfies $\partial^2 J/\partial s^2 = 0$:
\begin{align}
\frac{\partial^2 J}{\partial s^2} 
&= \int_0^T\int_0^T\frac{\partial\epsilon(s,t')}{\partial s}\mathcal{H}(t,t')\frac{\partial\epsilon(s,t)}{\partial s}\;dt\;dt' = \displaystyle\sum_{i=1}^{2\mathcal{N}}\sigma_i\left[\int_0^T\frac{\partial\epsilon(s,t)}{\partial s}\phi_i\;dt\right]^2 = 0. \label{eq:Js2}
\end{align}
Thus, we require that for some arbitrary, $L^2$-integrable function $f(s,t)$,
\begin{equation}
\frac{\partial\epsilon(s,t)}{\partial s} = f(s,t) -\displaystyle\sum_{l=1}^{2\mathcal{N}}\psi_l(t)\int_0^T{\psi_l(t')f(s,t')\;dt'},\;s\geq 0
\label{eq:epsilonevolve}
\end{equation}
at the global maximum, where the functions $\psi_l(t)$ are orthonormal (with respect to the $\|\cdot\|_T$ norm) with the same span as the functions $\phi_i(t)$ in Eq. (\ref{eq:hessianexpansion}).  The functions $\psi_l(t)$ may be obtained by applying the Gram-Schmidt process to the functions $\phi_i(t)$.  Equation (\ref{eq:epsilonevolve}) automatically satisfies $\partial^2 J/\partial s^2 = 0$, which is evident upon direct substitution into Eq. (\ref{eq:Js2}).  The differential equations in (\ref{eq:epsilonevolve}) are coupled to Hamilton's equations (\ref{eq:sysdym}) and initiated with an optimal field $\epsilon(0,t)$ already found to produce the global maximum value for $J$.

\section{Comparing the classical and quantum control landscapes} \label{sec:quantum}

Our analysis of the classical phase space trajectory control landscape has a mathematical analogue with the quantum state-to-state transition probability control landscape.\cite{rabitz2006topology,rabitz2004quantum} In particular, optimally controlling the $2N$-dimensional quantum wavefunction (i.e., its joint $N$ real and $N$ imaginary components) of a system with $N$ \emph{energy levels} is mathematically similar to optimally controlling the $2{\mathcal N}$ \emph{position and momentum state variables} of $n$ classical particles.  We emphasize that this analogy is not a direct physical comparison of the two classes of systems.  The latter distinction is evident, as controlling the motion of an $n$-particle quantum mechanical system requires a wavefunction which lies in an infinite-dimensional space.  Nevertheless, there are striking mathematical similarities between the $n$-particle classical and $N$-state quantum control landscapes, as these two cases describe analogous state-to-state ``transitions'' in their respective classical and quantum regimes.

The classical analogue of the quantum transition probability is the corresponding observable function $O({\bf z}(T)) = [{\bf z}(T)-{\bf z_{\rm tar}}]^T{\bf W}[{\bf z}(T)-{\bf z_{\rm tar}}]$, with ${\bf W}$ symmetric and negative-definite and ${\bf z_{\rm tar}}$ the target state.  Since $O({\bf z}(T))$ has only one critical point, at which ${\bf z}(T)= {\bf z_{\rm tar}}$ and $O({\bf z}(T))$ is globally maximized, the analysis in Section \ref{sec:eqns} shows that the classical target-state landscape is trap-free with no saddle points.  To obtain this result, we assumed that the functional {\color{black} mapping $\delta \epsilon(\cdot)\rightarrow\delta {\bf z}(T)$} is surjective, implying that appropriate perturbations $\delta\epsilon(t)$ in the control field can move the final state ${\bf z}(T)$ in an arbitrary direction $\delta{\bf z}(T)$.  Quantum mechanical studies make the analogous assumption that proper perturbations in the control field can enable the final state $|\psi(T)\rangle$ to vary in any direction $|\delta\psi(T)\rangle$.\cite{rabitz2006topology,rabitz2004quantum}

In order to show the analogue of the quantum mechanical case to the classical formulation in Section \ref{sec:eqns}, we summarize here the key quantum mechanical state-to-state landscape relations.  The objective functional for quantum mechanical state-to-state transition probability control may be written as $J_{{\rm qu}} \equiv \langle\psi(T)|f\rangle\langle f|\psi(T)\rangle$, where $|\psi(t)\rangle$ evolves from the initial state $|i\rangle$ at $t=0$ to the final state $|\psi(T)\rangle$.\cite{rabitz2006topology, rabitz2004quantum} The goal is to maximize $J_{\rm qu}$ (i.e., steer $|\psi(T)\rangle$ to be aligned with $|f\rangle$, up to an overall phase) in analogy with the classical goal of maximizing $J$ in Eq. (\ref{eq:cost}).  As with the critical point condition for the classical control landscape (Eq. (\ref{eq:chain})), the critical point of the quantum control landscape $J_{\rm qu}[\epsilon]$ corresponds to the condition that the gradient $\delta J_{{\rm qu}}/\delta\epsilon(t)$ be zero:
\begin{equation}\label{eq:quantumgradient}
\frac{\delta J_{\rm qu}}{\delta\epsilon(t)}=\left(\frac{\partial J_{\rm qu}}{\partial |\psi(T)\rangle},\frac{\delta|\psi(T)\rangle}{\delta\epsilon(t)}\right)=0,
\end{equation}
where we used the convenient notation $\left(w,v\right)={\bf Re}\left( w^\dagger v\right)$ to express the inner product of two
$N$-dimensional complex vectors.  Here ``{\bf Re}" is the real part of a complex number.

The evolution of the wave function $|\psi(t)\rangle$ follows the time-dependent Schr{\"o}dinger equation $(\partial/\partial t)|\psi(t)\rangle = -(\imath/\hbar)\big(H_0-\mu\epsilon(t)\big)|\psi(t)\rangle,\ |\psi(0)\rangle=|i\rangle$, where  $H_0$ and $\mu$ are, respectively, the field-free Hamiltonian and the dipole moment operator of the quantum system.  In order to avoid confusion with the classical Hamiltonian in Eq. (\ref{eq:hamiltonianfree}), we use explicitly different notation for the latter quantum mechanical operators.  For an $N$-level quantum system, due to (i) the preservation of the norm $\langle\psi(t)|\psi(t)\rangle=1$ and (ii) the
fact that the overall phase of $|\psi(T)\rangle$ is not physically relevant, the state $|\psi(t)\rangle$ may be considered as a column vector of up to $2N-2$ independent components.

The propagator $U(t,0)$ satisfies the equation $(\partial/\partial t)U(t,0) = -(\imath/\hbar)\left(H_0 - \mu\epsilon(t)\right)U(t,0)$ with $U(0,0) = I$, which is the quantum analogue of $M(t)$ in Eq. (\ref{eq:Mdym}).  We can solve for the functional derivative $\delta|\psi(T)\rangle/\delta\epsilon(t) = (\imath/\hbar)U(T,0)U^\dagger(t,0)\mu U(t,0)|\psi(0)\rangle$ and use the relation
$\partial J_{\rm qu}/\partial|\psi(T)\rangle=2|f\rangle\langle f|\psi(T)\rangle$ to obtain the
gradient
\begin{eqnarray}
\frac{\delta J_{\rm qu}}{\delta\epsilon(t)}
&=&\left(2U^\dagger(T,0)|f\rangle\langle f|\psi(T)\rangle,\frac{\imath}{\hbar} U^\dagger(t,0)\mu U(t,0)|\psi(0)\rangle\right), \label{eq:quantumje}
\end{eqnarray}
which leads to critical point controls when $J_{\rm qu}=0$ or $J_{\rm qu} = 1$, respectively, at the global minimum ($\langle f|\psi(T)\rangle=0$) or at the global maximum ($U^\dagger(T,0)|f\rangle\langle f|U(T,0)=|i\rangle\langle i|$). 
At both of these ``regular'' critical points, $\delta |\psi(T)\rangle/\delta\epsilon(t)$ is of maximum $2N-2$ rank.\cite{rabitz2006topology,rabitz2004quantum}

At the global maximum the corresponding Hessian can be
written as \cite{ho2006effective}
\begin{eqnarray}
{\cal H}_{\rm qu}(t,t')
&=&\frac{-2}{\hbar^2}{\bf Re}\left(\sum_{k\neq i}^N\langle i | \mu(t')|k\rangle\times\langle k|\mu(t)|i\rangle\right), \label{eq:quantumhessian}
\end{eqnarray}
which is negative-semidefinite and at most of rank $2N-2$.  The summation in Eq. (\ref{eq:quantumhessian}) is analogous to that of Eq. (\ref{eq:hessianexpansion}); both are sums of linearly independent functions, with the number of such functions equal to the rank of the derivative of the state with respect to the control.  Here we may associate $\sqrt{|\sigma_l|}\phi_l(t)$ in Eq. (\ref{eq:hessianexpansion}) with $\left\langle k|\mu(t)|i\right\rangle$ in Eq. (\ref{eq:quantumhessian}).  Thus, as \sout{with the classical circumstance}{\color{black} in the classical case}, the Hessian for the quantum state-to-state transition probability is symmetric and of finite rank when we assume surjectivity of the functional derivative $\delta|\psi(T)\rangle/\delta\epsilon(t)$.  In addition, the norm of $\delta J_{\rm qu}/\delta\epsilon(t)$ {\color{black} \sout{is }}and the trace of the quantum Hessian are bounded,\cite{rabitz2006topology} just as {\color{black} for \sout{are }}their classical counterparts (see Section \ref{sec:stability}).  Moreover, one may explore the level sets of quantum landscapes, much as Section \ref{sec:stability} considered the level set at the global maximum in the classical case.\cite{beltrani2011exploring}

\section{Numerical simulations} \label{sec:simulations}

The results in Section \ref{sec:eqns} on the existence of trap-free classical state-to-state control landscapes rest on some key assumptions, particularly the surjectivity of $\delta {\bf z}(T)/\delta\epsilon(t)$.  To assess the validity of these assumptions, we performed classical optimal control calculations based on Hamilton's equations for a range of model molecular systems with various Hamiltonians. {\color{black} Importantly, the success of these simulations bolsters the control landscape principles in Section \ref{sec:eqns}, but these optimal control studies are not meant to suggest that future experiments would operate with \emph{a priori} designed fields. As with systems operating in the analogous quantum dynamical regime, the discovery of optimal fields is likely best done directly in the laboratory with adaptive feedback control,\cite{brif2010control} especially for complex molecules. However, the establishment in this paper that classical control landscape principles share common behavior with their quantum mechanical counterparts provides a broad foundation to analyze and understand the dynamics of future molecular control experiments.}

In these calculations, we considered the objective functional
\begin{align}
J &= -\begin{bmatrix} \left({\bf q}(T)-{\bf q_{\rm tar}}\right) & \left({\bf p}(T)-{\bf p_{\rm tar}}\right) \end{bmatrix}^T{\color{black} {\bf K}}\begin{bmatrix} \left({\bf q}(T)-{\bf q_{\rm tar}}\right) \\ \left({\bf p}(T)-{\bf p_{\rm tar}}\right) \end{bmatrix} \nonumber \\ &= -({\bf q}(T)-{\bf q_{\rm tar}})^T{\bf K}_1({\bf q}(T)-{\bf q_{\rm tar}}) - ({\bf p}(T)-{\bf p_{\rm tar}})^T{\bf K}_2({\bf p}(T)-{\bf p_{\rm tar}}),
\label{eq:costused}
\end{align}
where $({\bf q}(T), {\bf p}(T))$ and $({\bf q_{\rm tar}},{\bf p_{\rm tar}})$ are pairs of real vectors denoting, respectively, the final and target states of the molecule in phase space.  {\color{black} The matrix ${\bf K} = {\rm diag}\left({\bf K}_1, {\bf K}_2\right)$, with ${\bf K}_1$ and ${\bf K}_2$ diagonal matrices $\mathcal{N}\times\mathcal{N}$ matrices scaling the position and momentum targets. The (positive) diagonal entries of ${\bf K}$ may be chosen for numerical convenience depending on the problem.} {\color{black} Since ${\bf K}$ represents the Hessian of the objective functional $J$ and is chosen to be positive-definite, we see from Section \ref{sec:hessian} that when ${\bf \delta z}(T)/\delta\epsilon(t)$ is surjective, $J$ has a unique critical point at its global maximum and the landscape is thus trap-free.}
The optimal control simulations were carried out using fourth-order symplectic integration of Hamilton's equations \cite{donnelly2005symplectic}
in conjunction with a gradient-based algorithm analogous to the D-MORPH method, which has been utilized in quantum optimal control simulations.\cite{rothman2006exploring,rothman2005observable} {\color{black} Many simulations were run, and a few cases will be shown to illustrate the principles of Section \ref{sec:eqns}.  We}\sout{This section} first present {\color{black} two cases for simple}\sout{a simple case illustrating the principles of Section \ref{sec:eqns} for} vibrational control of a heteronuclear diatomic molecule \sout{with the field }aligned along the \sout{molecular axis}{\color{black} field direction}.  We then present the results of several simulations with systems of coupled oscillators and $2n = 4$ position and momentum coordinates{\color{black} , as well as a chaotic system with a coupled quartic oscillator}.  In all of the tests, no traps on the landscape were encountered.

\subsection{Numerical techniques: Symplectic integration and D-MORPH optimization algorithms}\label{sec:numerical}

In the following simulations,  the initial trial control field $\epsilon_0(t)$ is chosen as a sum of four sine functions with randomly selected amplitudes and phases.  The {\color{black} control} field is allowed to freely vary as a discretized function of time, $0\leq t\leq T$, on an evenly spaced temporal grid, i.e., $t_j= j\Delta t\in[0,T]$, $j = 1,2,\ldots, \lfloor T/\Delta t\rfloor$, where $\Delta t$ is a small, fixed time-step.  Finding an optimal control field involves iterating two steps: (i) solving Hamilton's equations and (ii) updating the control field, until $J$ has reached a maximum value.

In step (i), a fourth-order symplectic integrator~\cite{candy1991symplectic} was used to {\color{black} solve} Hamilton's equations (\ref{eq:sysdym}) for the state evolution over the discretized time interval $[0,T]$.  A symplectic integrator was chosen for reasons of numerical stability, although other methods could be utilized.  The adopted method uses four substeps to evolve the state from time $t$ to time $t + \Delta t$.  Specifically, denoting the state at time $t$ by $({\bf q^0},{\bf p^0})$, for $i = 1,2,3,4$, we have
\begin{align}
{\bf q^i} &= {\bf q^{i-1}} + b_i\frac{\partial H}{\partial {\bf p}}({\bf p}^i)\Delta t, \label{eq:qint}\\
{\bf p^i} &= {\bf p^{i-1}} - c_i\frac{\partial H}{\partial {\bf q}}({\bf q}^{i-1})\Delta t\label{eq:pint}
\end{align}
with $({\bf q}(t+\Delta t),{\bf p}(t+\Delta t)) \equiv ({\bf q}^4,{\bf p}^4)$.  The values of the fixed parameters $b_i$ and $c_i$ are given in \cite{donnelly2005symplectic,mclachlan1992accuracy}.

In step (ii), the control field is refined using a gradient search algorithm based on D-MORPH \cite{rothman2005quantum} via a homotopy parameter $s$, such that the field is written as $\epsilon(s,t)$ and $\epsilon(0,t) = \epsilon_0(t)$. The D-MORPH equation to update the field is 
\begin{equation}
\frac{d\epsilon(s,t)}{ds} \equiv \beta\frac{\delta J}{\delta\epsilon(s,t)},\ \beta >0, \label{eq:epsilongrad}
\end{equation}
ensuring that
\begin{equation}
\frac{dJ}{ds} =\int_0^T\frac{\delta J}{\delta\epsilon(s,t)}\frac{d\epsilon(s,t)}{ds}\;dt
= \beta\int_0^T \left(\frac{\delta J}{\delta\epsilon(s,t)}\right)^2\;dt \geq 0.
\label{eq:Jgrad}
\end{equation}
Equation (\ref{eq:epsilongrad}) is discretized over time, as described above.  The field $\epsilon(s,t)$ evolves as $\epsilon(s,t)\rightarrow\epsilon(s+\Delta s,t)$, thereby increasing $J$. Integration of Eq. (\ref{eq:epsilongrad}) is stopped at the convergence criterion: {\color{black} $-J < 0.01$}.   Equation (\ref{eq:epsilongrad}) is integrated with respect to $s$ at each of the time points $t_j$ using a fourth-order Runge-Kutta method.  At each increment in $s$, the gradient $\delta J/\delta\epsilon(s,t)$ is computed with Eq. (\ref{eq:chain}) and $\delta J/\delta\epsilon(t) = \left[\partial J/\partial{\bf z}(T)\right]\left[\delta{\bf z}(T)/\delta\epsilon(s,t)\right]$ once Hamilton's equations are solved, using Eqs. (\ref{eq:qint}) and (\ref{eq:pint}) in step (i) based on the control field in the preceding iteration. 

\subsection{Example: Optimal control of a {\color{black} \sout{linear }}diatomic molecule}\label{sec:linearexample}

Consider a model vibrating diatomic molecule driven by a time-dependent control field $\epsilon(t),\ t\in[0,T]$, aligned with the molecular axis.  The diatom is modeled as a Morse oscillator with Hamiltonian
\begin{equation}
H = {\color{black} \frac{\overline{p}^2}{2}} + V(q)-D(q)\epsilon(t)= {\color{black} \frac{\overline{p}^2}{2}} + D_0\left(1-e^{-\alpha q}\right)^2 - Aqe^{-\xi q^4}\epsilon(t),
\label{eq:HF}
\end{equation}
where  $q$ and $\overline{p} = p/\sqrt{m}$ are, respectively, the relative distance and {\color{black} scaled} momentum between the two atoms.  The center of mass motion has been removed and the problem is expressed in internal coordinates; however, the control landscape in this formulation should still exhibit the general behavior identified in Section \ref{sec:eqns}.

{\color{black} The} parameters for the Morse potential $V(q)$ and dipole moment function $D(q)$ in Eq. (\ref{eq:HF}) are chosen to have the values: 
$m = 1732$,  $D_0 = 0.2101$, $\alpha = 1.22$, $A = 0.4541$ and $\xi = 0.0064$, 
all given in atomic units  (a.u.).\cite{efimov2003feedback,guldberg1991laser,stine1979classical} {\color{black} We define $\epsilon(t) = G(t)E(t)$, where $G(t) = \exp\left(-{\color{black} 10^{-4}(t - T/2)^2/2}\right)$ is a Gaussian envelope function, and then optimize with respect to $E(t)$.  The presence of the Gaussian guides the fields $\epsilon(t)$ towards the physically desirable behavior of being small at the beginning and end of the time interval $[0,T]$.  Two cases will be shown here, operating respectively in the strong and weak field regimes, to particularly illustrate that the principles in Section \ref{sec:eqns} are expected to be broadly applicable.

In the first case, t}he initial condition is chosen as  $(q_0,\overline{p}_0) = (0.2,0.0)$, corresponding to a slightly stretched bond (i.e., the equilibrium position is $q = 0$) with zero momentum.  The desired target state at $T = 320\pi$ is $(q_{{\rm tar}},\overline{p}_{\rm tar}) =(2.0,{\color{black} 0.014})$, which is far from the initial state.  {\color{black} Note that for the momentum, the distance between the initial value of 0.0 and the target of 0.014 is quite substantial, as we will demand precision at this level and the phase space excursion will range out to $\left|\overline{p}\right|\sim 0.8$.}  As the time interval is also relatively short, we expect that the optimal field will be strong.  For numerical stability, we take ${\bf K} = {\rm diag}\left(0.999, {\color{black} 1.732}\right)$ in Eq. (\ref{eq:costused}).

\begin{figure}
\centering
\includegraphics[width = 0.45\textwidth]{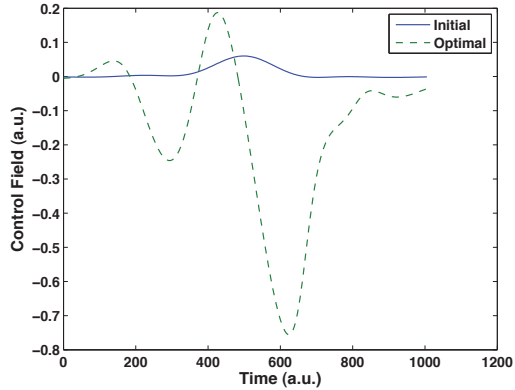}
\caption{Initial field and final optimal control for the single oscillator with a strong field.  The initial field yields the phase space point $(q(T), \overline{p}(T)) = (0.19, 3.6\times 10^{-4})$.  The target state was $(q_{\rm tar},\overline{p}_{\rm tar}) = (2.0, 0.014)$, and the optimal field achieved $(q(T),\overline{p}(T)) = (2.000, 0.015)$.}
\label{epsilon}
\end{figure}

Figure \ref{epsilon} shows the initial and final optimal control fields\sout{; no attempt was made to impose the constraint that the field dies out at $t = 0$ and $T$.  There is a general resemblance between the initial and final fields, which was seen in other simulations and commonly arises in quantum mechanical studies as well}.  The objective had the value $J = -3.25$ for the initial field, and the final field achieved the near optimal value $J = -1.87\times 10^{-8}$, corresponding to the final state $q(T) = 2.000$ and ${\color{black} \overline{p}(T) = 0.015}$.  The latter values are very close to the target state of (2.0, {\color{black} 0.014}).

\begin{figure}
\centering
\includegraphics[width = 0.55\textwidth]{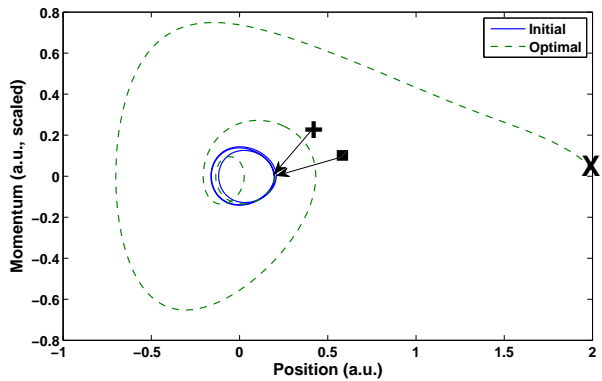}
\caption{Phase plane trajectories with the initial {\color{black} field} and final optimal control fields for the single oscillator.  The point denoted with $\times$ marks the {\color{black} final state} of the trajectory {\color{black} with the optimal field, while that denoted with + marks the final state with the initial field.  The \sout{gray }square $\blacksquare$ marks the initial point $\left(q_0, \overline{p}_0\right)$.}\sout{The inset amplifies the domain that exhibits an interesting feature in the phase plane trajectory with the optimal field.}}
\label{qp}
\end{figure}

Figure \ref{qp} shows the phase plane trajectories {\color{black} arising from\sout{for}} the initial {\color{black} field} and final optimal \sout{control }field.  \sout{The dynamics under the optimal field exhibit an interesting feature shown in the inset, evidently present to adjust the phase space trajectory to meet the strong disparity between $(q_0, p_0)$ and $(q_{\rm tar},p_{\rm tar})$.}The optimal control field substantially changes and ``stretches out'' the $q$ trajectory in order to reach $q_{\rm tar} = 2$ at the final time.  This behavior is consistent with the optimal field \sout{being very negative}{\color{black} having a large amplitude} towards the end of the trajectory, as shown in Fig. \ref{epsilon}\sout{\color{black} ; the final position changes dramatically towards the end of the trajectory, under the influence of this strong field.}.

{\color{black} A key feature of the analysis in Section \ref{sec:eqns} is the assumed surjectivity of $\delta {\bf z}(T)/\delta\epsilon(t)$.  For the illustration here, we calculate the singular values of the {\color{black} time-}discretized {\color{black} matrix} $\delta {\bf z}(T)/\delta\epsilon(t)$ to be 1.77 and 0.01, showing that the two functions $\delta q(T)/\delta\epsilon(t)$ and $\delta \overline{p}(T)/\delta\epsilon(t)$ are comfortably linear independent.}

\sout{A key feature of the analysis in Section \ref{sec:eqns} is the assumed surjectivity of $\delta {\bf z}(T)/\delta\epsilon(t)$.  For the illustration here, Fig. \ref{qpepsilon} shows $\delta q(T)/\delta\epsilon(t)$ and $\delta p(T)/\delta\epsilon(t)$, as functions of time $t$ for the optimal field. 
By inspection, these two functions are clearly linear independent, indicating that the $\left\{\delta {\bf z}(T)/\delta\epsilon(t)\right\}$ form a full-rank set of vectors.  Further numerical evaluations (not shown) demonstrate that $\left\{\delta {\bf z}(T)/\delta\epsilon(t)\right\}$ is also full-rank at non-optimal fields.}
\sout{The two trajectories in Figure \ref{qpepsilon} are not periodic, but each has a well-defined characteristic period of about 320 for $\delta q(T)/\delta\epsilon(t)$ and 360 for $\delta p(T)/\delta\epsilon(t)$.  Approximating the Morse potential $V(q)$ by a harmonic oscillator, the natural period of the system is 331, roughly matching the latter periods.}

{\color{black} We next show an example of the same diatomic model with a weak optimal field.  \sout{We now take t}{\color{black} T}he initial and target states {\color{black} are chosen} to be $\left(q_0, \overline{p}_0\right) = \left(-0.3, {\color{black} 0.012}\right)$ and $\left(q_{\rm tar}, \overline{p}_{\rm tar}\right) = \left(0.5, {\color{black} 0.12}\right)$, respectively, and {\color{black} we} substantially lengthen the control interval to $T = 3200\pi$.  The matrix ${\bf K}$ in Eq. (\ref{eq:costused}) is again taken to be ${\rm diag}\left(0.999, 1.732\right)$. {\color{black} Since we lengthen the control interval, we scale the Gaussian envelope to be $G(t) = \exp\left(-{\color{black} 10^{-6}(t - T/2)^2/2}\right)$.}

Figure \ref{fig:epsilonsweak} shows the optimal control field and resulting phase plane dynamics.  We see that the field is much weaker than that in Fig. \ref{epsilon}, with a maximum amplitude less than 0.06 a.u.  The phase plot dynamics are {\color{black} quasi-}periodic, and come quite close to the target state: with the optimal field, $\left(q(T), \overline{p}(T)\right) = \left(0.497, {\color{black} 0.119}\right)$, for an objective functional value $J = -8.46\times 10^{-6}$.  Figure \ref{fig:costweak} shows the corresponding evolution of the cost functional.

Finally, we examine the surjectivity of $\delta {\bf z}(T)/\delta\epsilon(t)$ at the optimal field.  For this illustration, Fig. \ref{fig:singular} shows the two functions $\delta q(T)/\delta\epsilon(t)$ and $\delta \overline{p}(T)/\delta\epsilon(t)$ with the optimal field.  We see that these are visibly linearly independent; the singular values of the {\color{black} time-}discretized $\delta {\bf z}(T)/\delta\epsilon(t)$ matrix are 3.978\sout{77} and 0.236\sout{57}, showing a comfortable linear independence.
\begin{figure}
\centering
\subfigure{\includegraphics[width = 0.45\textwidth]{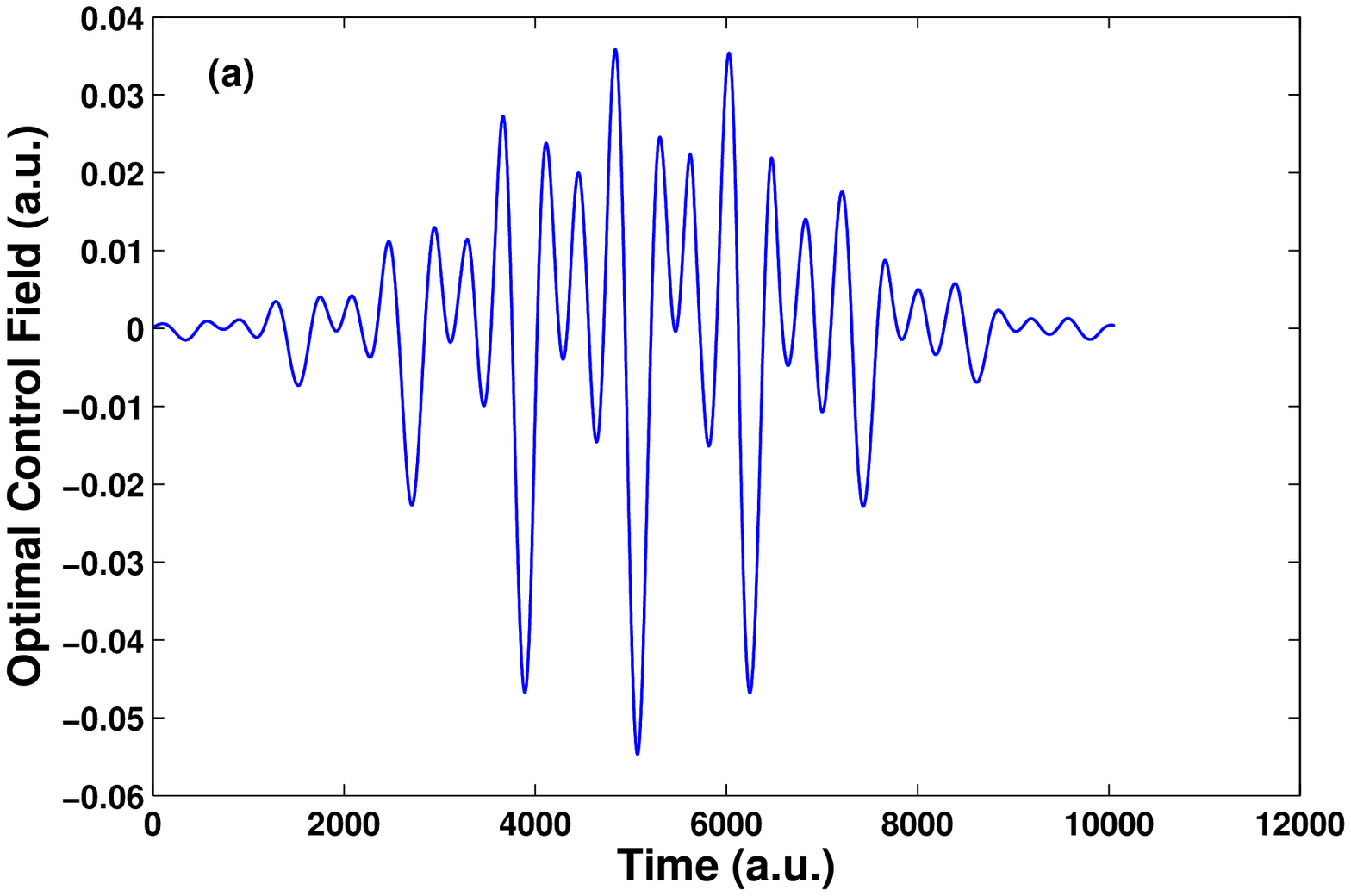}\label{fig:epsilonmove}}
\subfigure{\includegraphics[width = 0.48\textwidth]{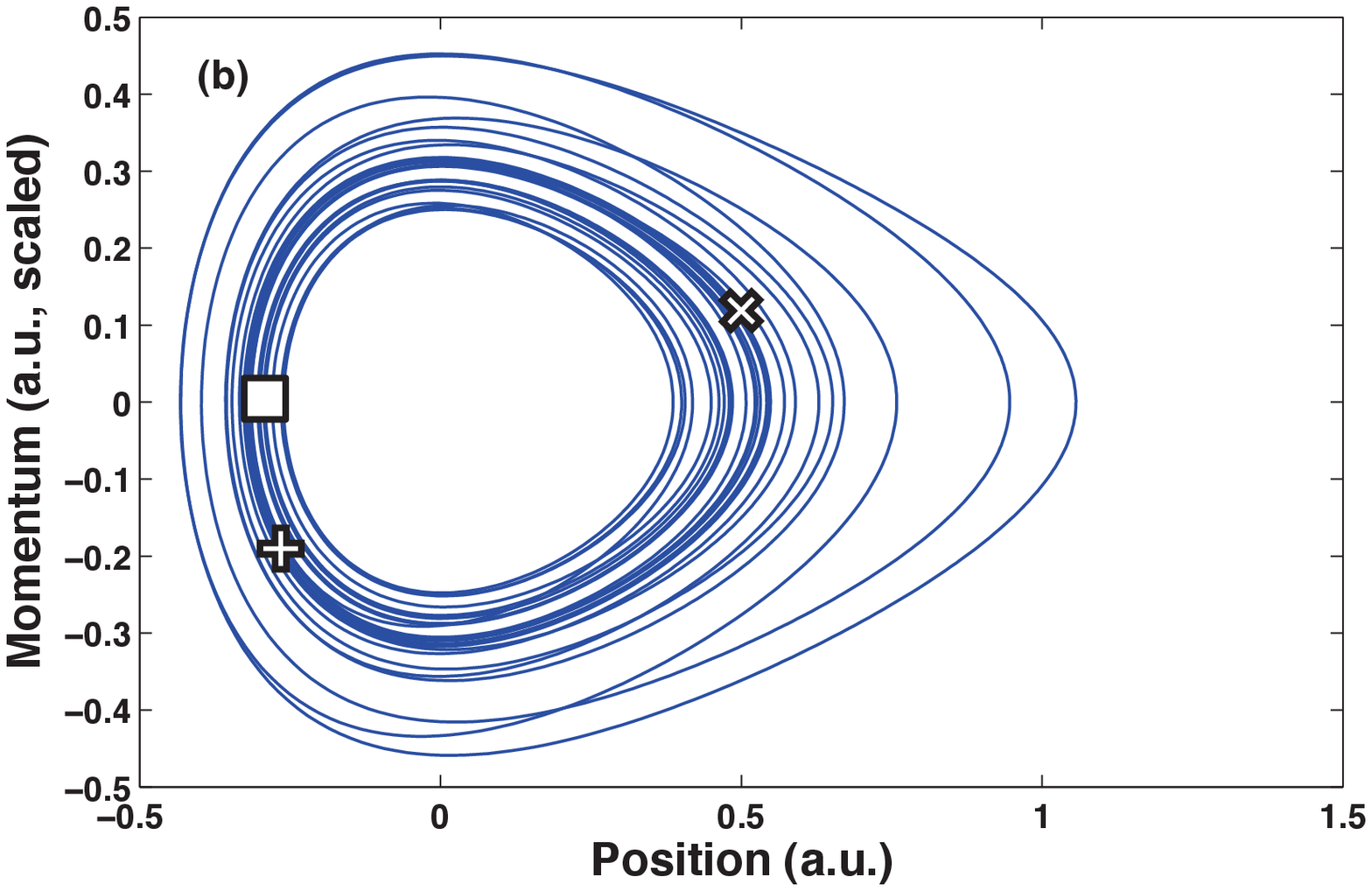}\label{fig:phasesmove}}
\caption{Optimal control field (a) and phase {\color{black} plane} plot (b) for the single oscillator with weak control field.  The `$\times$' marks the final point with the optimal field, and the `+' marks the final point with the initial field{\color{black} ; the \sout{gray }square $\square$ marks the initial point $\left(q_0, \overline{p}_0\right) = (-0.3, 0.012)$}.  With the initial field, the final state is $(q(T), \overline{p}(T)) = (-0.300, -{\color{black} 0.17{\color{black} 1}\sout{05}})$\sout{;} {\color{black} and} with the optimal field it is $(q(T), \overline{p}(T)) = (0.497\sout{4}, {\color{black} 0.119})$.  The target state is $(q_{\rm tar}, \overline{p}_{\rm tar}) = (0.5, {\color{black} 0.120\sout{1}})$.\sout{The momentum coordinate is scaled by $\sqrt{m}$.}}
\label{fig:epsilonsweak}
\end{figure}

\begin{figure}
\centering
\includegraphics[width = 0.5\textwidth]{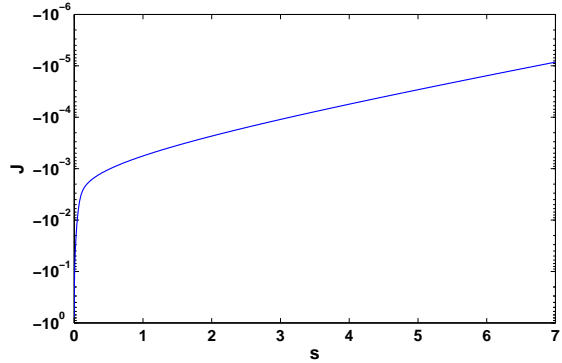}
\caption{Evolution of the objective functional $J$ versus {\color{black} $s$ for the single oscillator with {\color{black} a} weak control field}.  The optimization was stopped when $J = -8.46\times 10^{-6}$.}
\label{fig:costweak}
\end{figure}

\begin{figure}
\centering
\includegraphics[width = 0.55\textwidth]{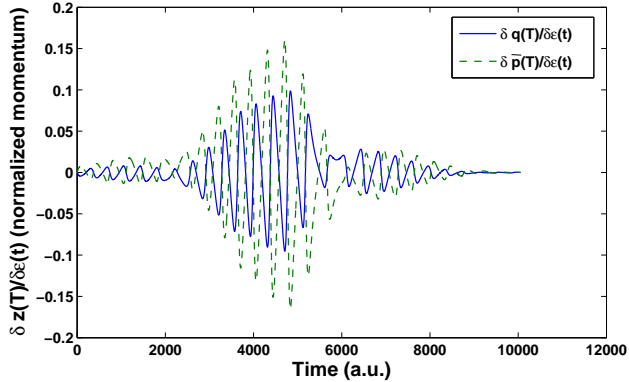}
\caption{\sout{Rep}{\color{black} P}resentation of $\delta{\bf z}(T)/\delta\epsilon(t)$ for the single oscillator with weak control field\sout{ with the momentum coordinate scaled by $\sqrt{m}$}.  The singular values are 3.97{\color{black} 8}\sout{77} and 0.23{\color{black} 6}\sout{57}{\color{black} , thereby showing clear linear independence of the two functions}.}
\label{fig:singular}
\end{figure}}
 
\subsection{Example: Optimal control of {\color{black} two} coupled Morse oscillators}\label{sec:example}

We consider a two-dimensional {\color{black} collinear} system of coupled Morse oscillators taken from previous works in the literature.\cite{sage1983energetics,radicioni1999application} We \sout{assume that}{\color{black} choose a coordinate frame where} the two oscillators are coupled through the potential and the momenta are decoupled.  The {\color{black} \sout{entire }}system is driven by a time-dependent control field $\epsilon(t),\ t\in[0,T]$.  We let $(q_1, q_2)$ and ${\color{black} (\overline{p}_1, \overline{p}_2) = (p_1/\sqrt{m_1}, p_2/\sqrt{m_2})}$ denote the two position and {\color{black} scaled} momentum coordinates respectively.  The Hamiltonian is given by
\begin{align}
H = &\frac{1}{2}\left({\color{black} \overline{p}_1^2 + \overline{p}_2^2}\right) + V(q_1,q_2)-D(q_1,q_2)\epsilon(t) \nonumber \\ = &{\color{black} \frac{\overline{p}_1^2}{2} + \frac{\overline{p}_2^2}{2}} + D_0\left[\left(1 - e^{-\alpha_1 q_1}\right)^2 + \left(1 - e^{-\alpha_2 q_2}\right)^2 +2\beta \left(1 - e^{-\alpha_1 q_1}\right)\left(1 - e^{-\alpha_2 q_2}\right)\right] \nonumber \\ &- \left(q_1e^{-\xi q_1} - q_2e^{-\xi q_2}\right)\epsilon(t).
\label{eq:SO2}
\end{align}
The center of mass motion has been removed and the problem is expressed in internal coordinates; as in the single oscillator example, the control landscape in this formulation should still exhibit the general behavior identified in Section \ref{sec:eqns}.

The parameters for the \sout{Morse }potential $V(q{\color{black} _1, q_2})$ and dipole moment function $D(q{\color{black} _1, q_2})$ in Eq. (\ref{eq:SO2}) are chosen to have the values: 
$D_0 = 0.2102$, $\alpha_1 = 1.1282$, $\alpha_2 = 0.9712$, $\beta = 0.1$, and $\xi = 0.1$,
all given in atomic units  (a.u.).\cite{sage1983energetics}  The masses are chosen as $m_1 = 10.0$, $m_2 = 5.0$.  Sixteen distinct simulations were performed with the initial conditions for $q_1$, $q_2$, $\overline{p}_1$, $\overline{p}_2$ chosen randomly at intervals of spacing 0.2 between $-1$ and 1 {\color{black} for the position coordinates; the momentum coordinates $\overline{p}_1$ and $\overline{p}_2$ were chosen at intervals of $0.06$ between $-0.32$ and $0.32$ and $0.09$ between $-0.45$ and $0.45$ respectively}.  The equilibrium position of the field-free Hamiltonian is at $q_1 = q_2 = 0$.  The desired target state at $T = 20\pi$ is $(q_{{\rm tar},1},q_{{\rm tar},2},\overline{p}_{{\rm tar},1},\overline{p}_{{\rm tar},2}) = (2.0, 0.5, 0, -{\color{black} 0.22})$.

\begin{figure}
\centering
\includegraphics[width = 0.94\textwidth]{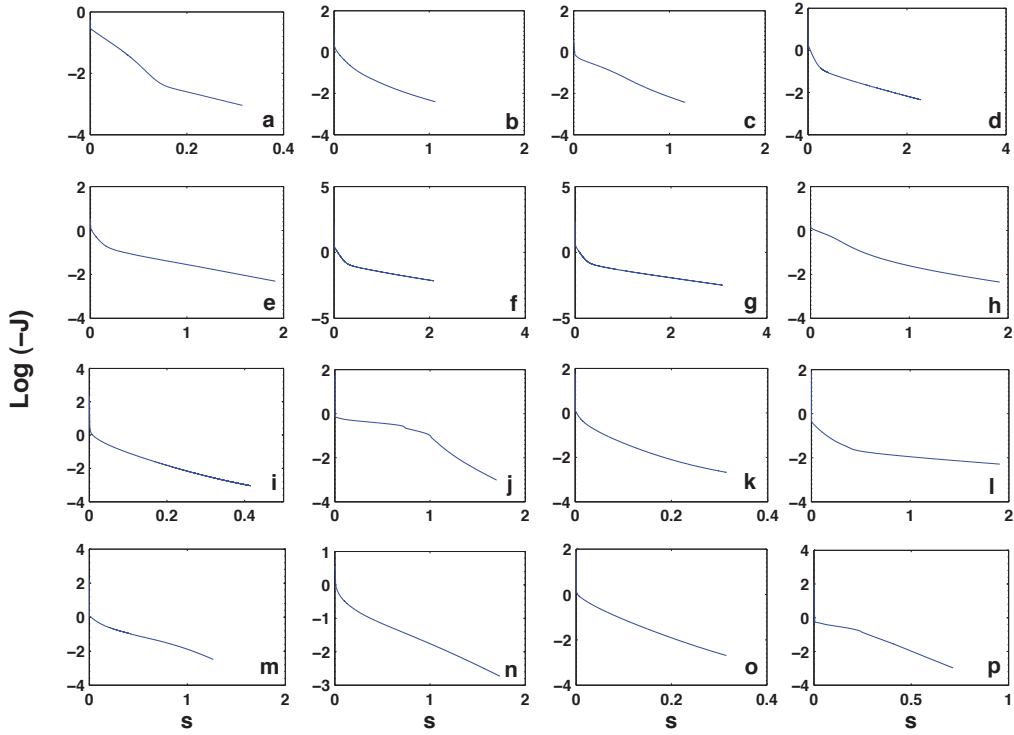}
\vspace{-0.2in}
\caption[Evolution of the objective functional versus the generalized iteration index.]{\label{cost_2d} The evolution of the objective functional $J$ versus $s$\sout{, which can be viewed as a generalized iteration index,} for sixteen different simulations using the two coupled oscillators.  The objective functional value approached zero in all cases, and the optimization was stopped when the values reached those shown in Table \ref{obj}.}
\end{figure}

Figure \ref{cost_2d} shows the increasing objective (i.e., {\color{black} $\log(-J)$}) as $\epsilon(s,t)$ evolves with $s$ for each of the sixteen simulations{\color{black} ; here $J$ is given by Eq. (\ref{eq:costused}) with ${\bf K} = {\rm diag} \left(1, 1, 0.1, 0.2\right)$}.  The initial control field was chosen as a sum of four sine functions with random amplitudes between 0 and 1 and frequencies at {\color{black} approximately} 0.5, 1.0, 1.5, and 2.0 times the frequencies of the $(q_1, \overline{p}_1)$ {\color{black} and $(q_2,\overline{p}_2)$} field-free motion.  {\color{black} The frequencies of} {\color{black} \sout{these systems}the two oscillators} {\color{black} are close to each other due to the choices of $\alpha_1$ and $\alpha_2$ in the Hamiltonian (\ref{eq:SO2}).}  The increase in $J$ generally is very rapid at the beginning and slows down as $s$ becomes large, due to the diminishing gradient $\delta J/\delta\epsilon(s,t)$.  In some cases (e.g., simulations j and p), the rate of increase is initially slow, and then increases.  These occurrences likely correspond to the evolution starting out at areas in the landscape of lower slope.  Nevertheless, the increase in the objective functional was strictly monotonic in all cases with no evidence of traps.  No attempt was made to traverse the level set at the maximum of $J$ as discussed in Section \ref{sec:eqns}.  The final state coordinates at the optimal control field are shown in Table \ref{obj}.

\begin{table}
\centering
\begin{tabular}{|c|cc|cc|cc|cc|}
\hline
Simulation & $q_1(0)$ & $q_1(T)$ & $q_2(0)$ & $q_2(T)$ & ${\color{black} \overline{p}}_1(0)$ & ${\color{black} \overline{p}}_1(T)$ & ${\color{black} \overline{p}}_2(0)$ & ${\color{black} \overline{p}}_2(T)$ \\ \hline
a & -0.2 & 1.982 & 0.2 & 0.504 & -{\color{black} 0.19} & {\color{black} 0.007} &  -{\color{black} 0.18} & -{\color{black} 0.225} \\
b & 0.8 & 1.958 & 0.8 & 0.495 & {\color{black} 0.19} & {\color{black} 0.013} &  -{\color{black} 0.27} & -{\color{black} 0.213} \\
c & 0.2 & 1.967 &  0 & 0.491 & {\color{black} 0.13} & {\color{black} 0.013} &  {\color{black} 0.18} & -{\color{black} 0.211} \\
d & 0.4 & 1.957 & 1.0 &  0.493 & 0 & {\color{black} 0.013} & -{\color{black} 0.18} & -{\color{black} 0.210} \\
e & -0.2 & 1.956 & 0 & 0.493 & 0 & {\color{black} 0.014} &  {\color{black} 0.36} & -{\color{black} 0.209} \\
f & 0.4 & 1.948 & -1.0 & 0.491 & -{\color{black} 0.19} &  {\color{black} 0.016} & -{\color{black} 0.27} & -{\color{black} 0.207} \\
g & 1.0 & 1.964 & -1.0 & 0.495 & 0 & {\color{black} 0.011} & -{\color{black} 0.09} & -{\color{black} 0.212} \\
h & 0 & 1.962 & -0.8 &  0.491 & -{\color{black} 0.13} & {\color{black} 0.014} & {\color{black} 0.09} & -{\color{black} 0.209} \\
i & -0.8 & 2.018 & 0.2 & 0.498 & {\color{black} 0.13} & -{\color{black} 0.008} & {\color{black} 0.09} & -{\color{black} 0.223} \\
j & -0.6 & 2.015 & -0.8 & 0.508 & {\color{black} 0.06} & -{\color{black} 0.007} & -{\color{black} 0.09} & -{\color{black} 0.230} \\
k & -0.6 & 2.025 & 0.2 & 0.500 & {\color{black} 0.06} & -{\color{black} 0.012} & 0 & -{\color{black} 0.225} \\
l & 0 & 1.957 & 0.4 & 0.494 & 0 & {\color{black} 0.014} & {\color{black} 0.18} & -{\color{black} 0.208} \\
m & -0.2 & 2.019 & 0.6 & 0.511 & {\color{black} 0.06} & -{\color{black} 0.009} & {\color{black} 0.27} & -{\color{black} 0.244} \\
n & 0.8 & 1.976 & 0 & 0.494 & -{\color{black} 0.06} & {\color{black} 0.009} & {\color{black} 0.18} & -{\color{black} 0.215} \\
o & -0.6 & 2.023 & -0.2 & 0.489 & {\color{black} 0.13} & -{\color{black} 0.011} & -{\color{black} 0.18} & -{\color{black} 0.218} \\
p & -0.6 & 2.016 & -0.6 & 0.503 &  {\color{black} 0.25} & -{\color{black} 0.009} & -{\color{black} 0.09} & -{\color{black} 0.225} \\ \hline

\end{tabular}
\caption{Initial state conditions and final state coordinates with the optimal fields of the simulations shown in Fig. \ref{cost_2d}.}
\label{obj}
\end{table}

\begin{figure}
\centering
\includegraphics[width = 0.55\textwidth]{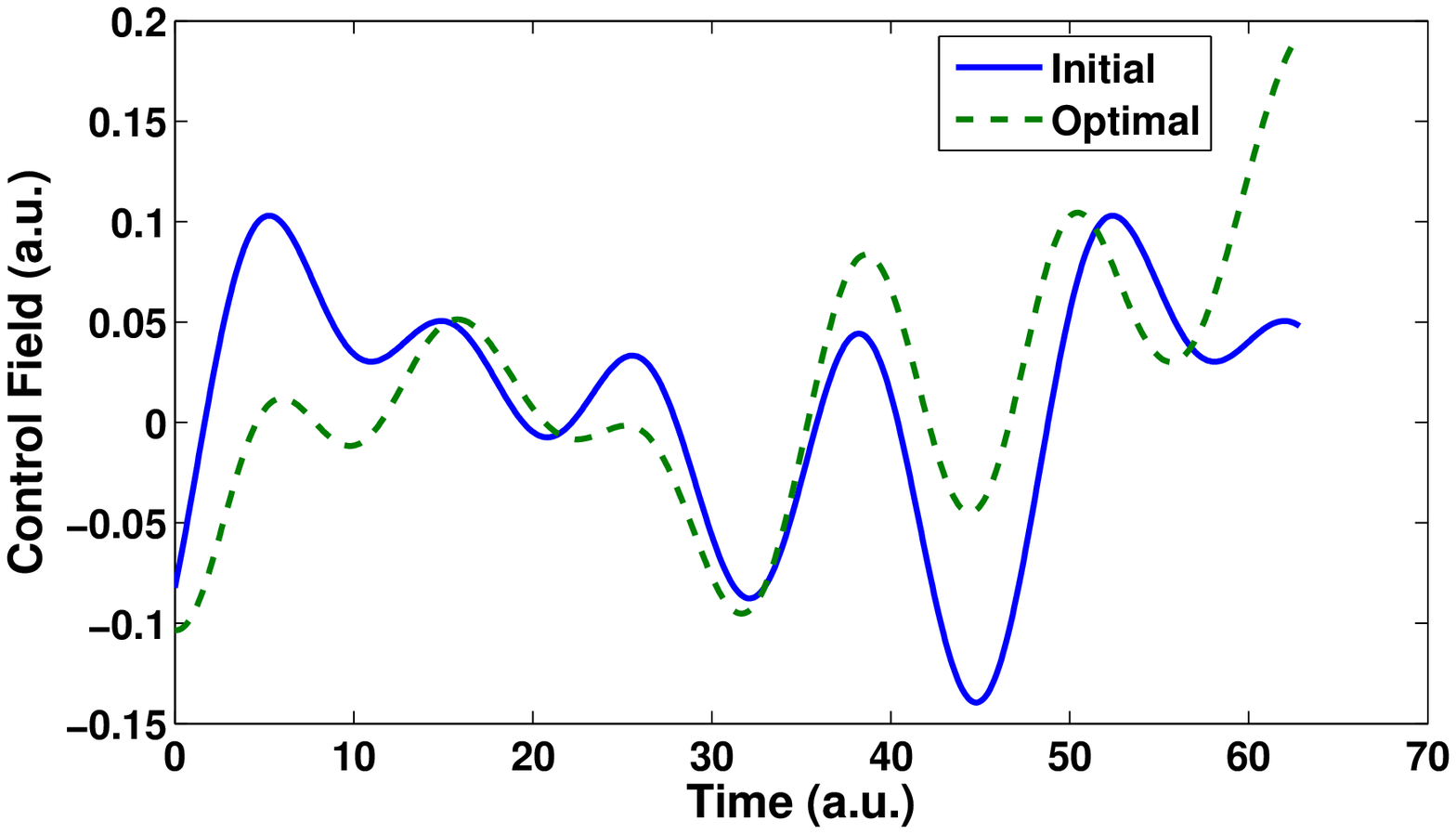}
\caption[Initial and final optimal control fields for simulation {\color{black} case o} with the two oscillators.]{\label{epsilon_2d} Initial and final optimal control fields for simulation {\color{black} case o} with the two coupled oscillators.  The initial field yields the state coordinates $(q_1(T), q_2(T), \overline{p}_1(T), \overline{p}_2(T))$ = (10.61, -0.40, {\color{black} 0.52, 0.23}).  The target state was $(q_{{\rm tar},1},q_{{\rm tar},2},{\color{black} \overline{p}_{{\rm tar},1},\overline{p}_{{\rm tar},2}) = (2.0, 0.5, 0, -0.22})$, and the optimal field achieved $(q_1(T),q_2(T),\overline{p}_1(T),\overline{p}_2(T))$ = (2.02, 0.49, {\color{black} -0.013, -0.22}).}
\end{figure}

\begin{figure}
\subfigure{\includegraphics[width = 0.45\textwidth]{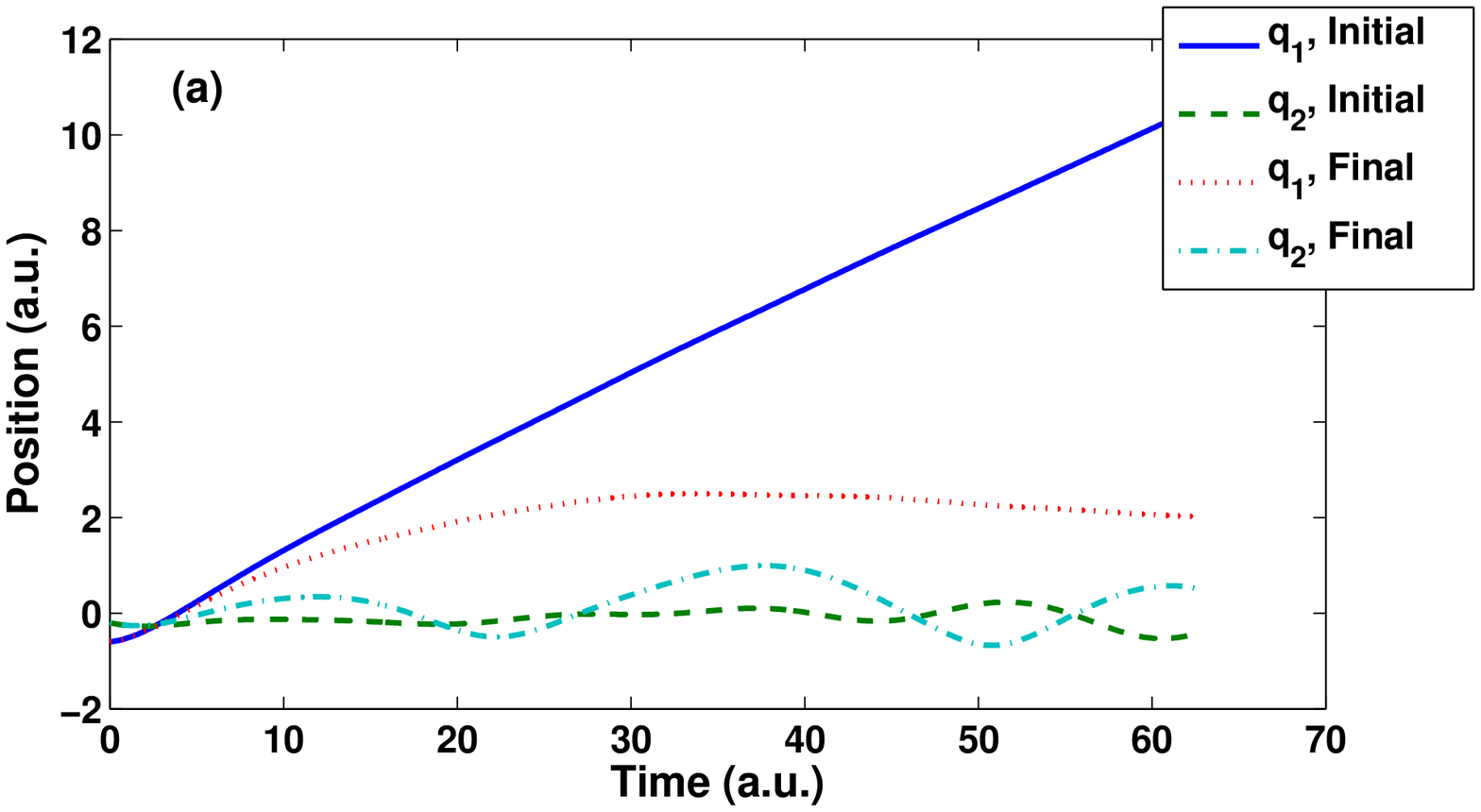}}
\subfigure{\includegraphics[width = 0.45\textwidth]{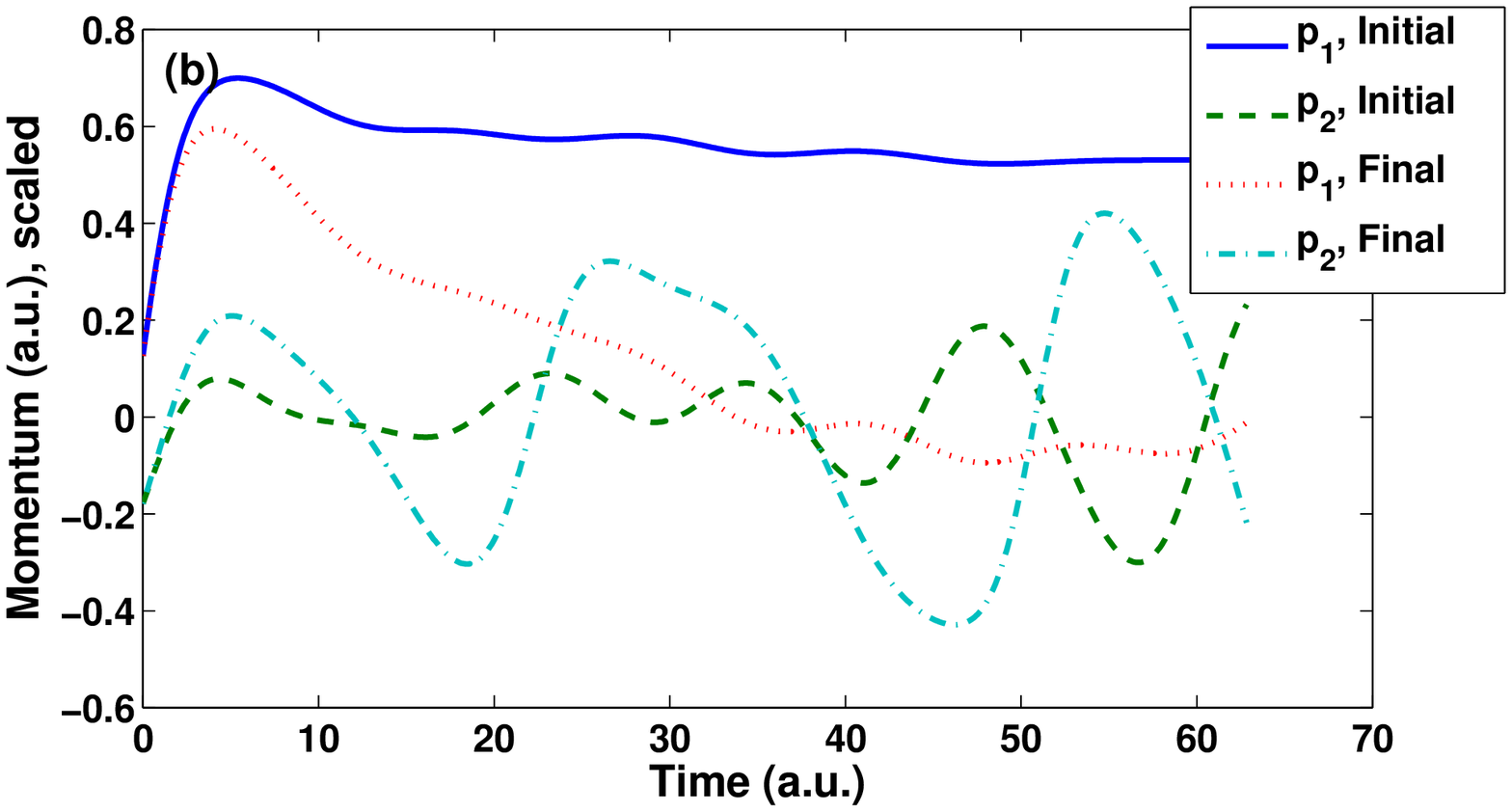}}
\caption{\label{qandp_2d} Position (a) and momentum (b) trajectories for {\color{black} case o of} the two coupled oscillators with initial and final control fields shown in Fig. \ref{epsilon_2d}.}
\end{figure}

\begin{figure}
\centering
\includegraphics[width = 0.55\textwidth]{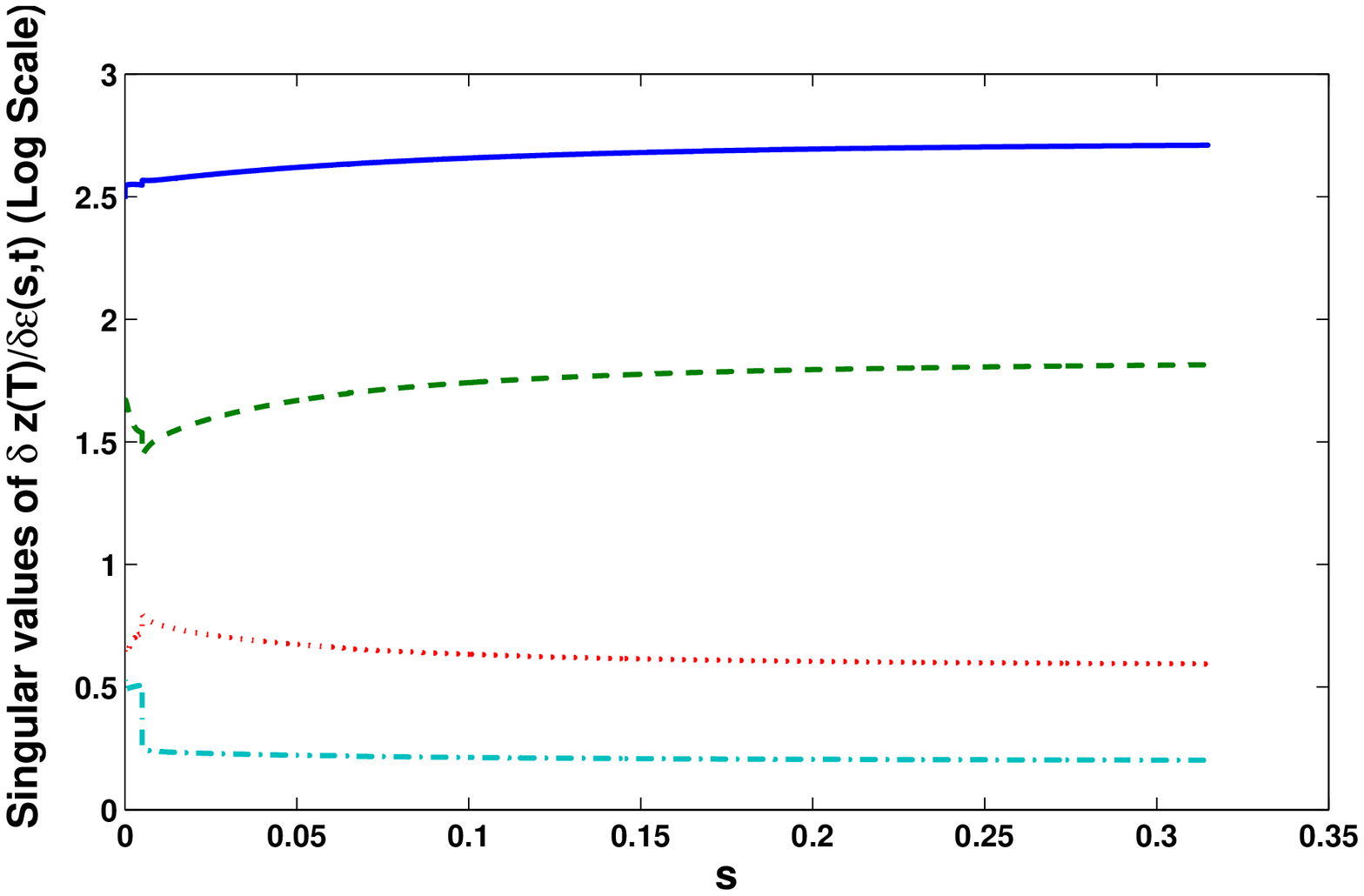}
\caption[Singular values of the derivatives of the final state with respect to the final optimal control field for the two coupled oscillators.]{\label{singular_2d} The singular values of $\delta {\bf z}(T)/\delta\epsilon(s,t)$ as a function of $s$ over the control field evolution to climb the landscape for {\color{black} case o of} the two coupled oscillators.  The finite, nonzero singular values are consistent with the surjectivity assumption in the analysis of Section \ref{sec:eqns}.}
\end{figure}

Figure \ref{epsilon_2d} shows the initial {\color{black} field and final optimal control field} for the simulation o; no attempt was made to impose the constraint that the field die out at $t = 0$ and $T$.  The moderate difference between the initial and final fields in Fig. \ref{epsilon_2d} has a dramatic impact on the {\color{black} final achieved state.}  The objective had the value $J = -78.71$ for the initial field, and the final field achieved the much higher and near optimal value $J = -0.00{\color{black} 19}$, corresponding to the final state $q_1(T) = 2.02$, $q_2(T) = 0.49$, ${\color{black} \overline{p}_1(T) = -0.013}$ and and ${\overline{p}_2(T) = -0.22}$.  The latter values are very close to the target state of (2.0, 0.5, 0, {\color{black} -0.22}).  Figure \ref{qandp_2d} shows the state trajectories for the initial and final optimal control fields in this example trajectory.  The first position coordinate $q_1(T)$ with the initial field has a value \sout{at time $T$ }far above the target of 2.0.  The optimal control field substantially changes and ``compresses'' this $q_1$ trajectory in order to reach $q_1(T) = 2.02$ at the final time.  Similar behavior was found for the other position and momenta coordinates.  The assumed surjectivity of $\delta {\bf z}(T)/\delta\epsilon(s,t)$ was also checked, and Fig. \ref{singular_2d} shows the eigenvalues of the {\color{black} time-}discretized $\delta {\bf z}(T)/\delta\epsilon(s,t)$ matrix as the control field $\epsilon$ evolves with $s$. 
The eigenvalues show comfortable linear independence of the \sout{columns}{\color{black} rows} of $\delta{\bf z}(T)/\delta\epsilon(s,t)$ for all $s$ values.  Similar results were observed for the other examples shown in Fig. \ref{cost_2d}.

\subsection{Example: Optimal Control of a Chaotic {\color{black} Coupled Quartic Oscillator}}

{\color{black} We next} consider \sout{a numerical example of }optimal control of a {\color{black} single particle of unit mass in a} coupled quartic oscillator \sout{system }{\color{black} potential. The dynamics are followed for sufficient time to {\color{black} exhibit characteristics of chaos ($T = 20\pi/3$), while staying within a numerically reliable regime}.\cite{schwieters1991optimal}}  The Hamiltonian of this system {\color{black} is}
\begin{equation}
H = \frac{1}{2}\left(p_1^2 + p_2^2 + q_1^4 + q_2^4\right) - kq_1^2q_2^2 - q_1\epsilon(t),
\label{eq:chaoticH}
\end{equation}
where $(q_1,q_2)$ and $(p_1, p_2)$ {\color{black} denote} respectively the position and momentum of the two degrees of freedom system.  As in the {\color{black} simulation for a diatomic molecule (Figs. \ref{epsilon} and \ref{fig:epsilonsweak}), the field is defined as} $\epsilon(t) = G(t)E(t)$, with $G(t) {\color{black} = \exp\left(-{\color{black} 10^{-5}(t - T/2)^2/2}\right)}$ a Gaussian envelope function. {\color{black} We choose} $k = 0.6$, {\color{black} such that} all of the energy phase space exhibits chaotic behavior.\cite{meyer1986theory} The initial condition is chosen as $(q_1,q_2) = (-0.4, 0.6)$ and $(p_1, p_2) = (0.2, 0.5)$.  The desired target state is $(q_{1,{\rm tar}},q_{2,{\rm tar}}) = (0.5, -0.5)$ and $(p_{1,{\rm tar}}, p_{2,{\rm tar}}) = (0.3, 0)$.  {\color{black} Achieving control is especially challenging, since only $q_1$ is directly influenced by the control field as indicated in Eq. \ref{eq:chaoticH}. We take ${\bf K}$ to be the identity matrix in the definition of the objective functional (\ref{eq:costused}).}

\begin{figure}
\centering
\subfigure{\includegraphics[width = 0.48\textwidth]{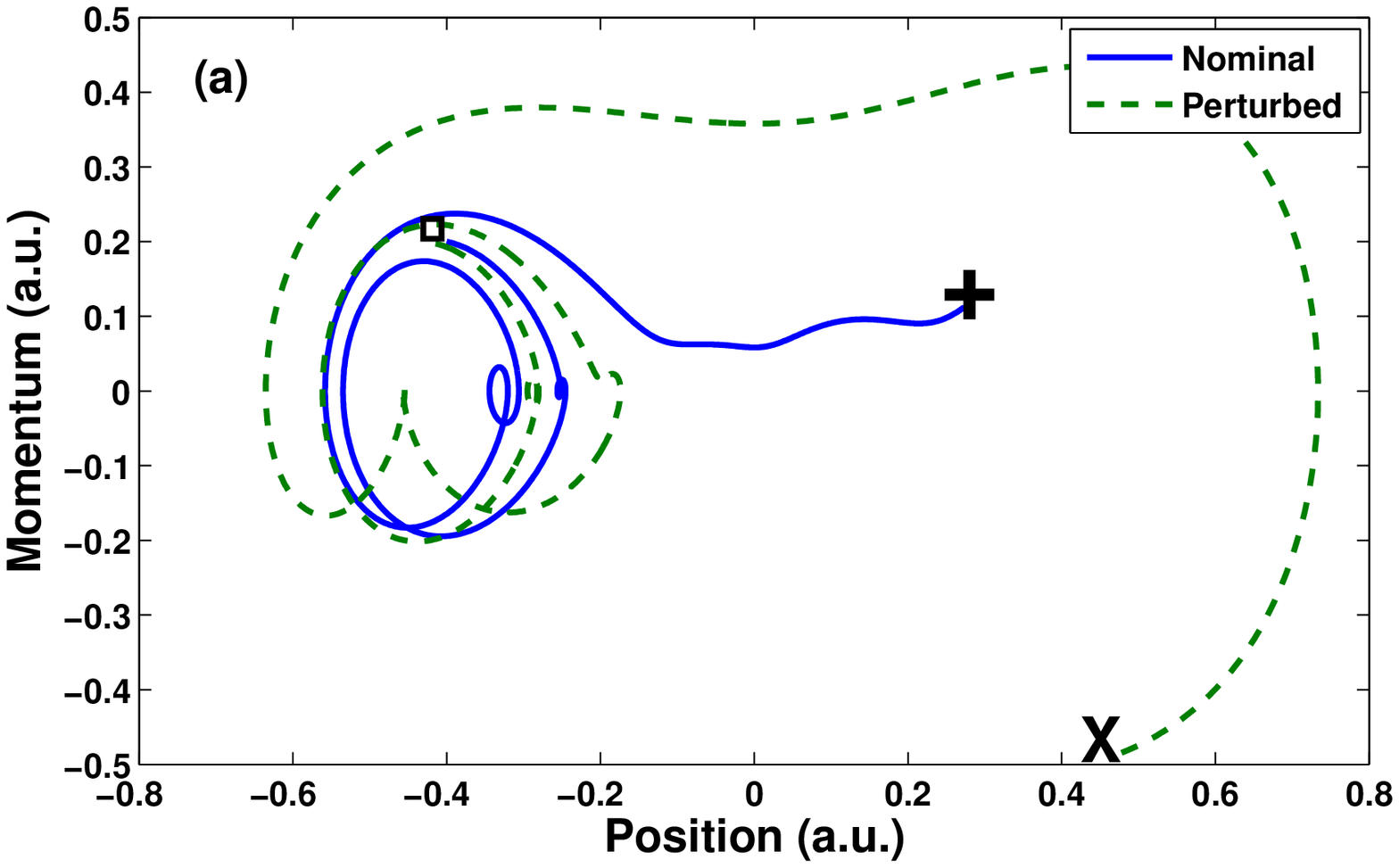}}
\subfigure{\includegraphics[width = 0.48\textwidth]{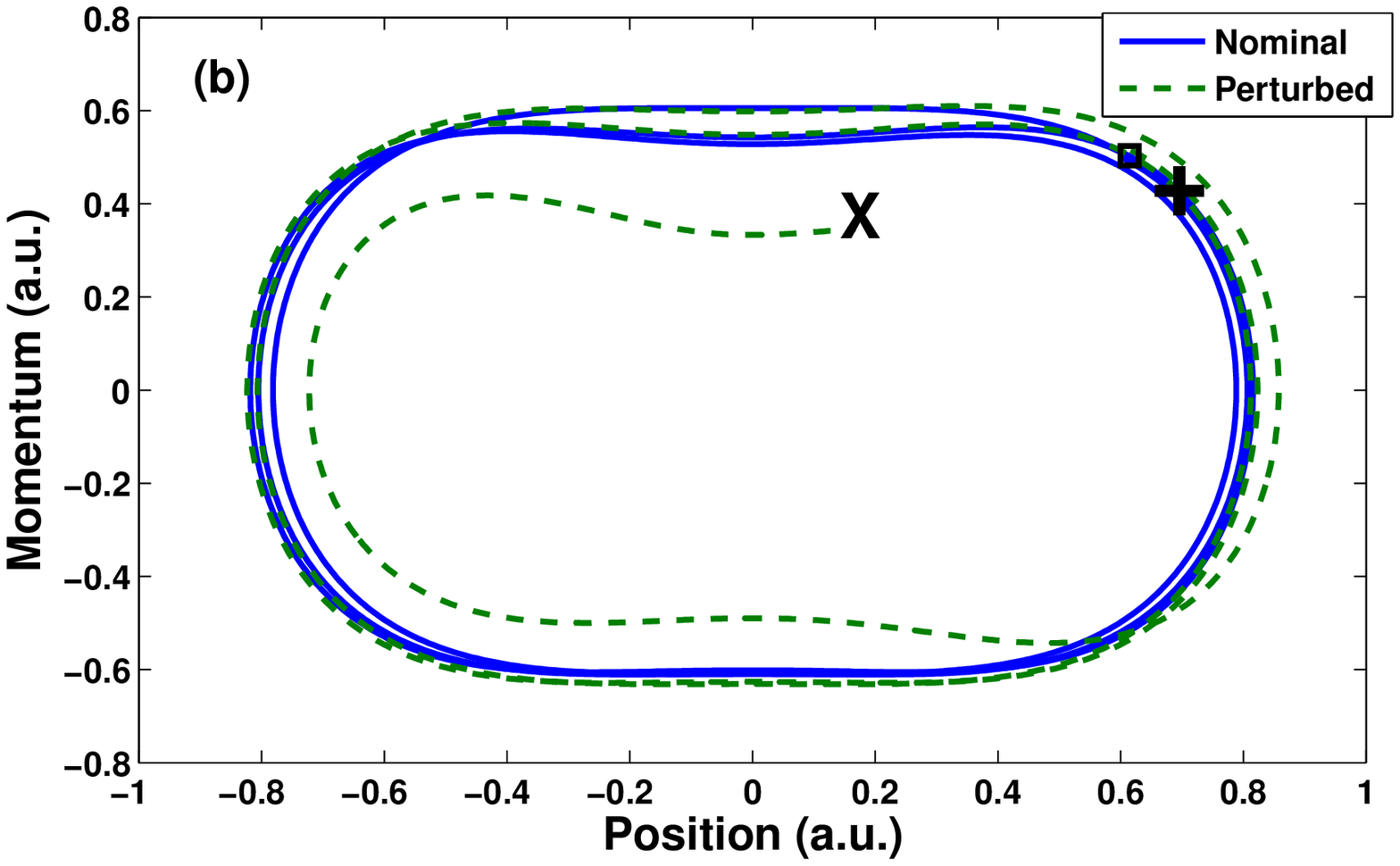}}
\caption{Phase-space illustrations {\color{black} (a) for $(q_1, p_1)$ and (b) for $(q_2, p_2)$} of the chaotic system's response to perturbations in the initial state values with no control field (nominal initial state: $\left(q_1(0), q_2(0), p_1(0), p_2(0)\right) = (-0.4, 0.6, 0.2, 0.5)$; perturbed initial state{\color{black} :} $\left(q_1(0), q_2(0), p_1(0), p_2(0)\right) = (-0.414, 0.617, 0.197, 0.515)$\sout{(-0.3927, 0.5994, 0.2071, 0.4980)}). A square marks the (approximate) location of the initial states in the phase plots; a '+' marks the final state of the nominal trajectories, and a 'X' marks the final state of the perturbed trajectory.}
\label{fig:chaos_none}
\end{figure}

\begin{figure}
\centering
\subfigure{\includegraphics[width = 0.48\textwidth]{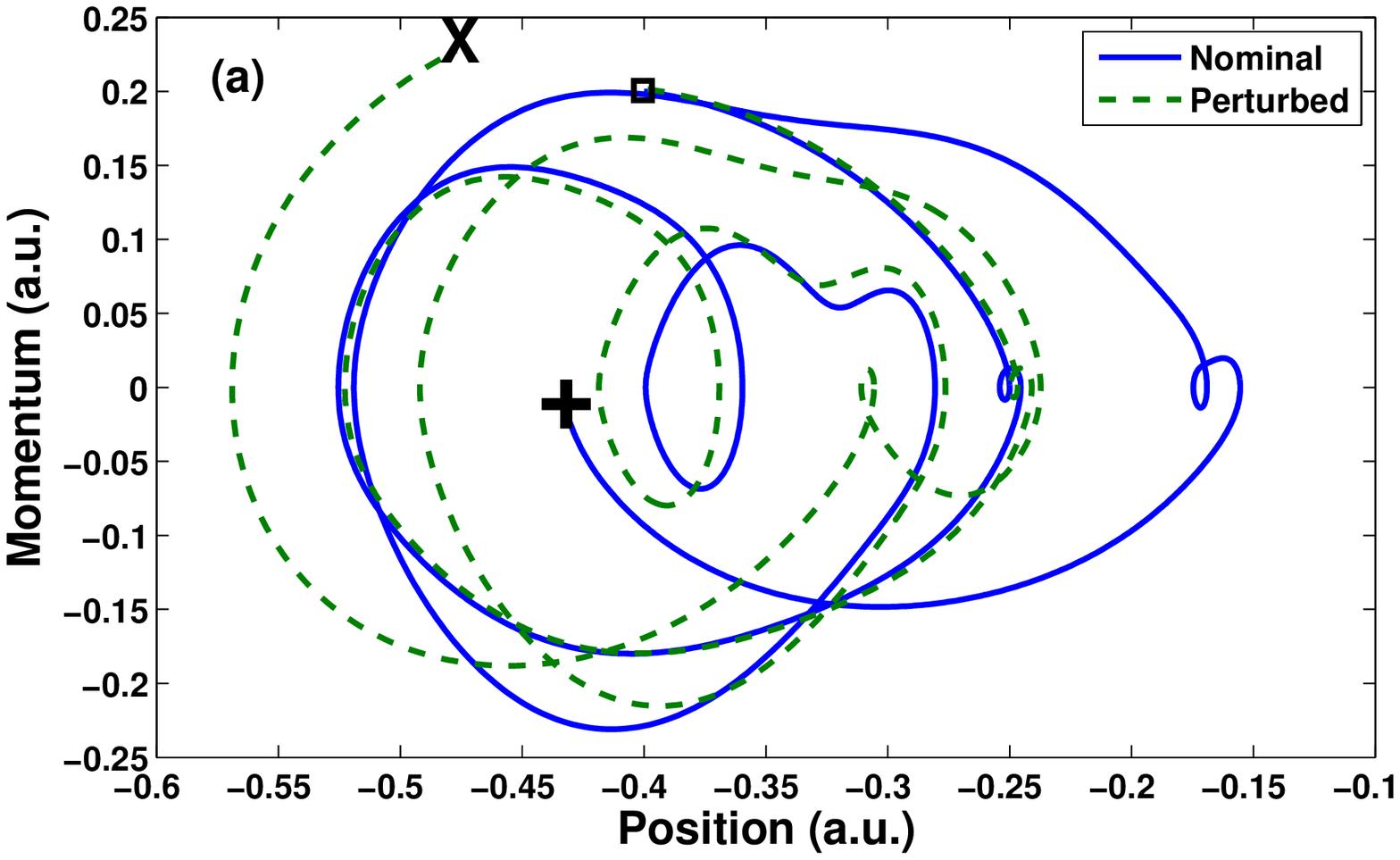}}
\subfigure{\includegraphics[width = 0.48\textwidth]{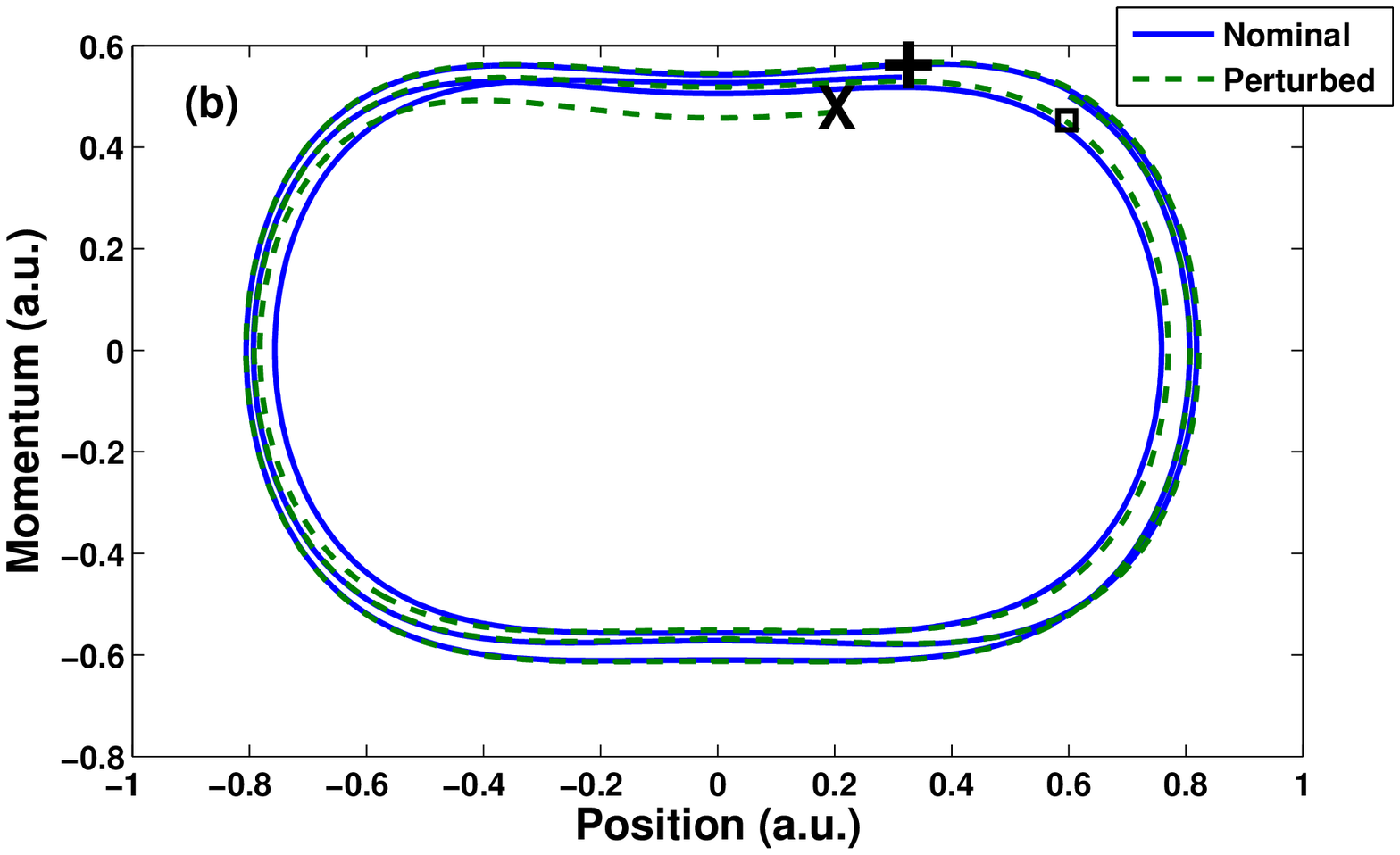}}
\caption{\label{fig:chaos_field} Phase-space illustrations {\color{black} (a) for $(q_1, p_1)$ and (b) for $(q_2, p_2)$} of the chaotic system's response to perturbations in the initial state values with {\color{black} the initial} control field present (nominal initial state{\color{black} :} $\left(q_1(0), q_2(0), p_1(0), p_2(0)\right) = (-0.4, 0.6, 0.2, 0.5)$; perturbed initial state{\color{black} :} $\left(q_1(0), q_2(0), p_1(0), p_2(0)\right) = (-0.3985, 0.6014, 0.2014, 0.5007)$). A square marks the (approximate) location of the initial states in the phase plots; a '+' marks the final state of the nominal trajectories, and a 'X' marks the final state of the perturbed trajectory.}
\end{figure}

\begin{figure}
\centering
\includegraphics[width = 0.55\textwidth]{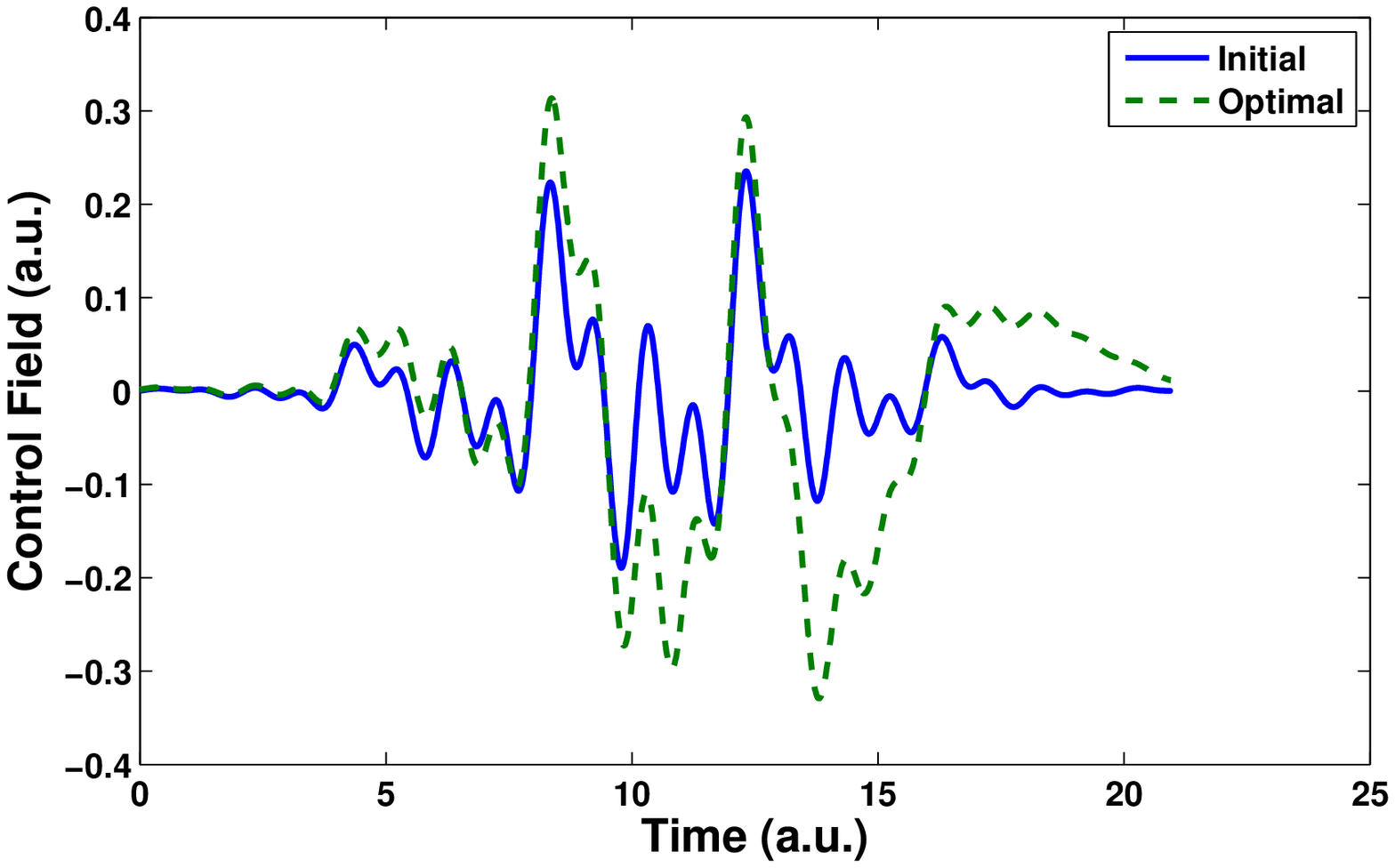}
\caption[Initial and final optimal control fields.]{\label{epsilon_diff} Initial and final optimal control fields.  The initial field yields the phase space point $(q_1(T), q_2(T), p_1(T), p_2(T)) = (-0.43, 0.31, -0.02, 0.54)$.  The target state was $(q_{1,{\rm tar}},q_{2,{\rm tar}}, p_{1,{\rm tar}}, p_{2,{\rm tar}}) = (0.5, -0.5, 0.3, 0)$, and the optimal field achieved $(q_1(T), q_2(T), p_1(T), p_2(T)) = (0.480, -0.593,  0.277,  0.013)$.}
\end{figure}

Figures \ref{fig:chaos_none} and \ref{fig:chaos_field} show the response of the phase-space trajectories to small perturbations in initial conditions, with no control field (Fig. \ref{fig:chaos_none}) and with the presence of the initial control field (Fig. \ref{fig:chaos_field}), i.e., the solid curve in Fig. \ref{epsilon_diff}. In these examples, the state variables at the final time take on very different values, despite the small initial perturbation {\color{black} (especially so in Fig. \ref{fig:chaos_field} with the field present)}. We note that the $(q_1, p_1)$ trajectory is somewhat more affected than the $(q_2, p_2)$ trajectory due to the larger values of $q_2$ and $p_2$ relative to $q_1$ and $p_1$ and the symmetry of the Hamiltonian without control field (Eq. \ref{eq:chaoticH}) with respect to $(q_1, p_1)$ and $(q_2, p_2)$. {\color{black} Additionally, the higher sensitivity in Fig. \ref{fig:chaos_field}(a) over that of \ref{fig:chaos_field}(b) is likely due to the field term in the Hamiltonian depending on the $q_1$ coordinate.} {\color{black} The} behavior {\color{black} in Figs. \ref{fig:chaos_none} and \ref{fig:chaos_field}} reflects the chaotic nature of this system, even at the modest final time. {\color{black} Similar perturbation studies of the single and coupled Morse oscillators {\color{black} in Sections \ref{sec:linearexample} and \ref{sec:example}} revealed much less mixing than observed for the quartic oscillator, as well as smaller discrepancies in the state variables at the final time.}

{\color{black} Figure \ref{epsilon_diff} shows th{\color{black} at both the} initial and final (near-optimal) control fields possess roughly the same periodicity. Both the initial and final control fields decay to a value near zero at $t = 0$ and $t = T$ due to the Gaussian envelope.}  The objective has the value $J = -1.92$ at the initial control field, which decreased to $J = -0.0097$ at the final control field, corresponding to a final state of $(q_1(T), q_2(T), p_1(T), p_2(T)) = (0.480, -0.593,  0.277,  0.013)$.  This achieved final state is quite close to the target of $(q_{1,{\rm tar}},q_{2,{\rm tar}}, p_{1,{\rm tar}}, p_{2,{\rm tar}}) = (0.5, -0.5, 0.3, 0)$.
\begin{figure}
\centering
\includegraphics[width = 0.55\textwidth]{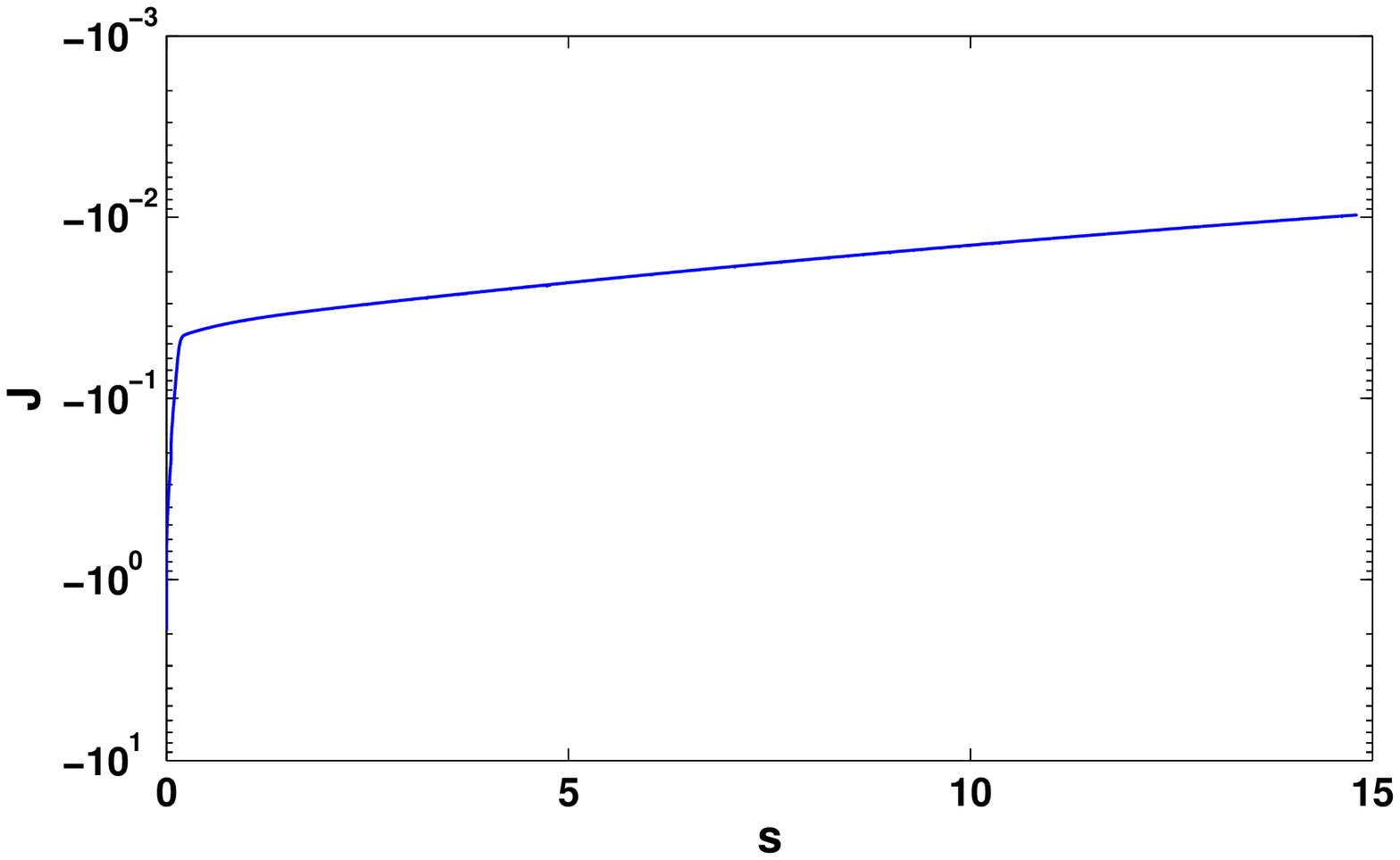}
\caption[Evolution of the objective functional versus the generalized iteration index.]{\label{cost_diff} The evolution of the objective functional $J$ versus {\color{black} $s$.}  The  objective continued to increase monotonically and the process was stopped when $J = -0.0097$ was achieved at $s = 14.8$.}
\end{figure}

Figure \ref{cost_diff} shows the increasing objective as {\color{black} a function of} $s$.  The objective increases very rapidly at the beginning of the trajectory and then increases more slowly, approaching {\color{black} zero}.  The increase was strictly monotonic, with no evidence of any traps encountered.

\begin{figure}
\subfigure{\includegraphics[width = 0.48\textwidth]{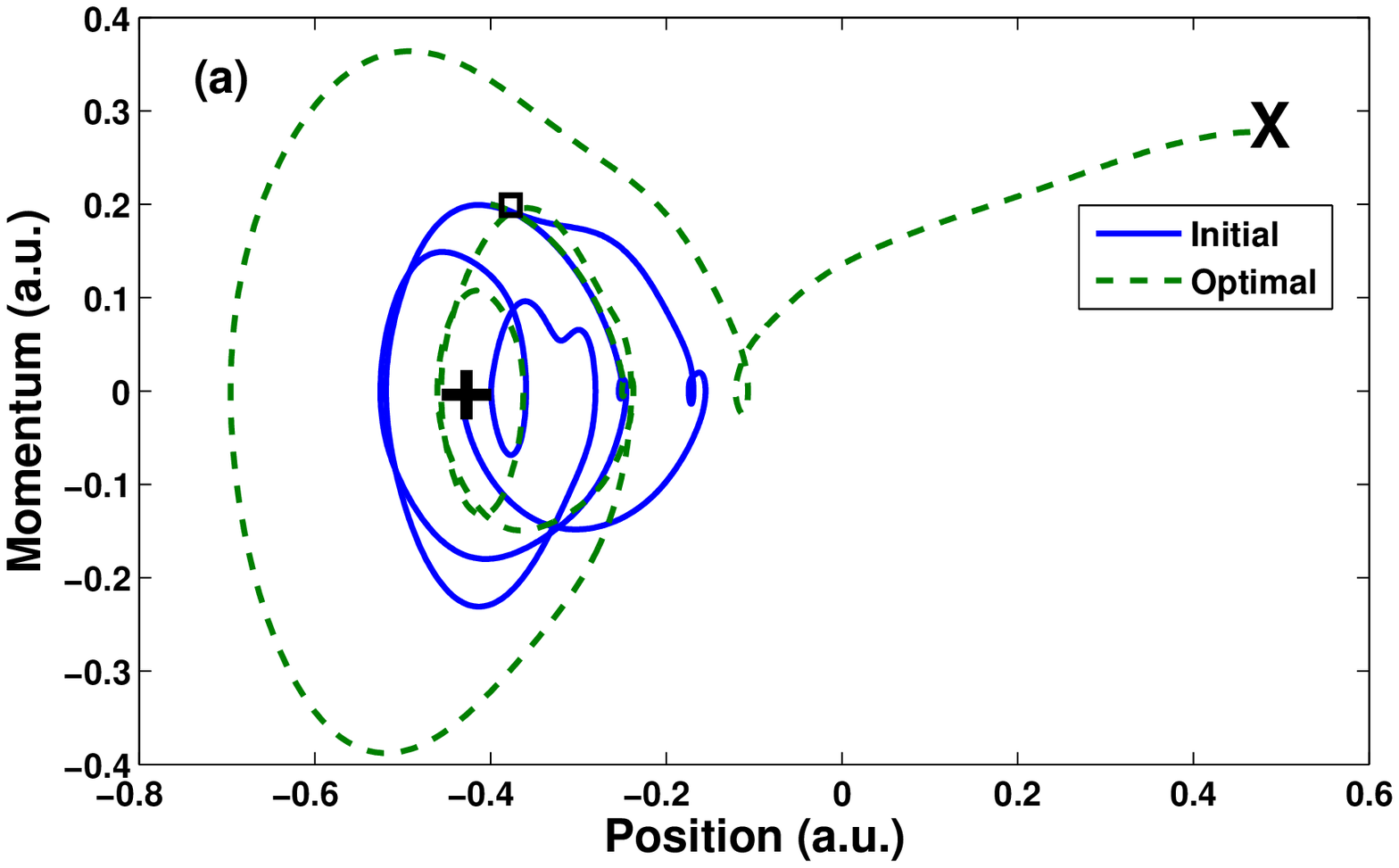}}
\subfigure{\includegraphics[width = 0.48\textwidth]{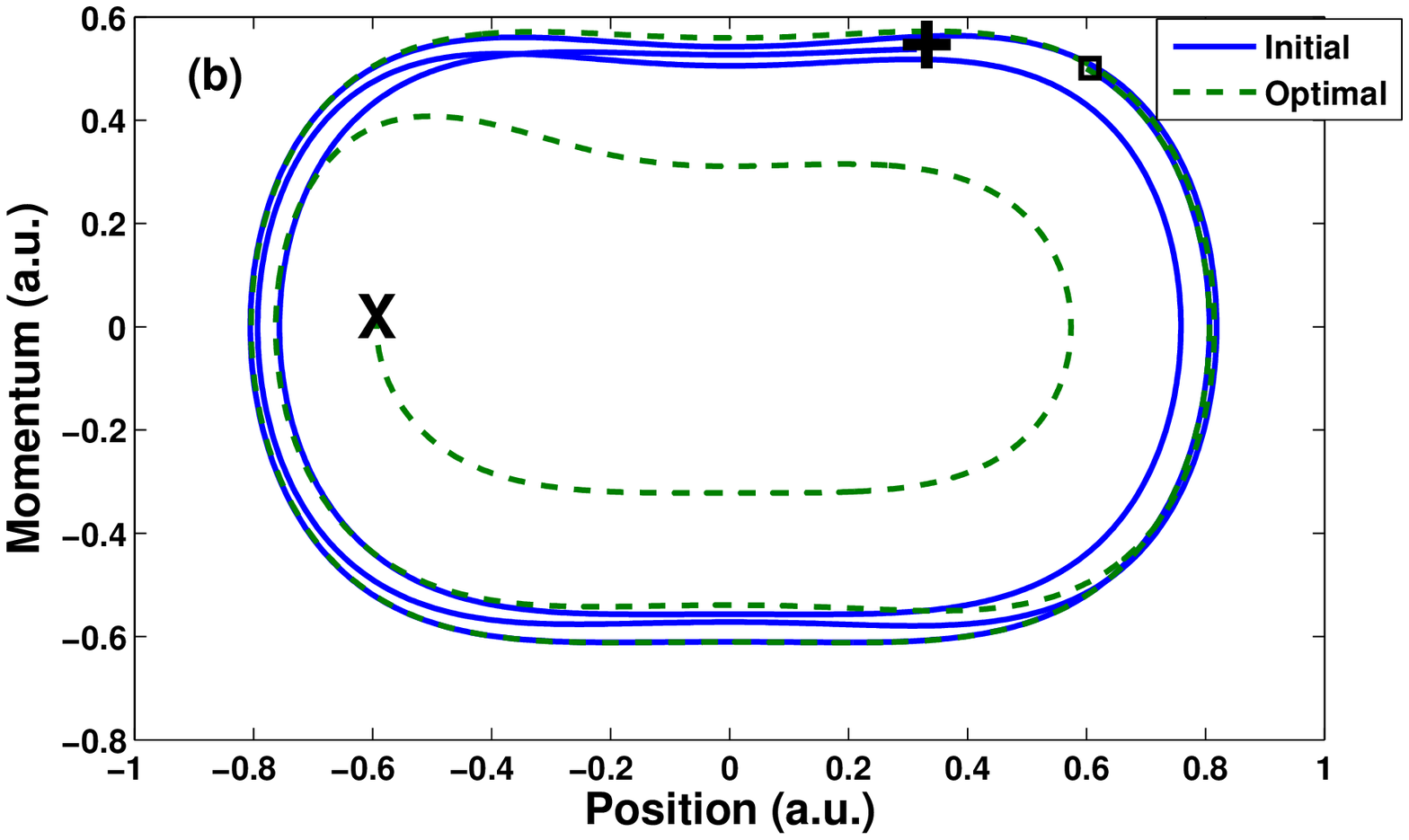}}
\caption{\label{qandp} Phase {\color{black} plane} plots of the position and momentum{\color{black} , $(q_1, p_1)$ in (a) and $(q_2, p_2)$ in (b),} trajectories with initial and optimal control fields. The final points are marked with an `+' and `X' for the initial and final fields{\color{black} ,} respectively; the initial point at $t = 0$ is marked with a square.}
\end{figure}

Figure \ref{qandp} depicts the evolution of the phase space coordinates $(q_1, q_2, p_1, p_2)$ with the initial and optimal control fields.  All four coordinates oscillate with rough periodicity under the influence of either control field{\color{black} , as shown by the cyclical nature of the phase space trajectories}.  {\color{black} The optimal control field substantially redirects} the position and momentum evolution in order to achieve the desired target states\sout{. Though the optimal control field remains relatively small at the end of the control interval $[0, T]$}, with the greatest distortion coming near the final time.

Finally, we find that the rank of $\delta {\bf z}(T)/\delta\epsilon(s,t)$ for each value of $s$ is 4, indicating that $\left\{\delta {\bf z}(T)/\delta\epsilon(s,t)|\;t\in[0,T]\right\}$ is surjective throughout the control field evolution.  Since the objective functional $J$ in Eq. \ref{eq:costused} has no critical points except at the global maximum, this finding {\color{black} is consistent with} no traps {\color{black} being encountered.} For other systems that exhibit more chaotic behavior and may be harder to control numerically, we may consider an ensemble average over several initial conditions.\cite{schwieters1991optimal}

Further simulations of the {\color{black} three} systems considered above, as well as other multi- and single-particle systems, showed no evidence of traps.  The $\delta{\bf z}(T)/\delta\epsilon(t)$ trajectories were also linear{\color{black} ly} independent, implying that $\delta{\bf z}(T)/\delta\epsilon(t)$ is full-rank or surjective.  This {\color{black} collective numerical evidence} supports the assumption that this behavior is likely the norm for physically realistic systems.

\section{Conclusion} \label{sec:conclusion}

This paper examines optimal control of a single phase space trajectory for a system following Hamiltonian classical dynamics. We have shown that for these systems, all critical points of the optimal control landscape $J[\epsilon(t)]$ are critical points of the observable $O({\bf z}(T))$, under specified assumptions including surjectivity of {\color{black} the functional mapping $\delta\epsilon(\cdot)\rightarrow\delta{\bf z}(T)$} \sout{$\delta{\bf z}(T)/\delta\epsilon(t)$} and boundedness of the potential function over the sampled phase space.  In particular, if the observable is a quadratic function aiming to reach a target state, a gradient-based search starting from an arbitrary field $\epsilon_0(t)$ should reach a critical (optimal) field $\epsilon(t)$ achieving this target state.  Moreover, the Hessian at the optimal critical point has an infinite-dimensional nullspace, indicating an inherent degree of robustness to control field noise. {\color{black} The rank of the Hessian may be used to reveal the effective number of particles of a complex molecule participating in the controlled dynamics; concomitantly, the control field structure addressing these particles and their mutual interactions may be identified by examining the eigenvectors of the Hessian.} The classical mechanical phase space state-to-state control results found here {\color{black} are \sout{have }}clear analogues with the goal of maximizing quantum mechanical pure state transition probabilities.  Interestingly, this analogy lies in correlating the number of classical particles with the number of quantum mechanical states.  Further study is {\color{black} needed} to seek additional classical-quantum control {\color{black} analogies \sout{analogues }}and relationships {\color{black} to fully understand control behavior in this bridged dynamical regime}.

Our findings extend previous studies \cite{pechen2010unified} from \emph{open} classical systems to \emph{closed}, single trajectory systems.  This prior work employed a kinematic analysis,\cite{pechen2010unified} while the present paper started with the dynamical equations.  The resulting analysis in Sections \ref{sec:criticalpoints} and \ref{sec:hessian} rested on the nature of the objective functional and the regularity of the controls considered.  These results have important fundamental implications for the control of classical mechanical systems with Hamiltonian dynamics, and in particular provide a firm foundation for controlled molecular dynamics of possibly complex polyatomic systems.\cite{rapaport2004art}

Numerical simulations confirmed the surjectivity of $\delta{\bf z}(T)/\delta\epsilon(t)$, and no evidence was found for traps on the landscape in these studies.  In practical molecular control applications, likely only a portion ${\bf z}'(t)$ of the state vector ${\bf z}(t)$ {\color{black} would \sout{will }}be specified for control, implying that there {\color{black} would \sout{will }}be a reduced demand on surjectivity for only $\delta {\bf z}'(T)/\delta\epsilon(t)$.  However, so-called singular controls, at which this set of derivatives is not surjective, also exist for both classical and quantum systems.\cite{bonnard2003singular,wu2012singularities} Such critical points are not yet well understood in the context of either classical or quantum optimal control, calling for additional analysis.  Moreover, this paper examines only the control of single trajectory systems.  Therefore, a natural next step to consider is that of controlling an ensemble of trajectories, with the state described by a probability distribution in the phase space in analogy with ensemble-based control of quantum systems.\cite{rabitz2006optimal} Collectively, these studies may form a basis for considering control behavior that transcends the classical and quantum regimes.

\section*{Acknowledgements}

This work was supported in part by the US Department of Energy and the Princeton Plasma Physics Laboratory.  CJ-W acknowledges the NDSEG Fellowship.  RBW acknowledges support from NSFC (Nos. 60904034 and 61134008).

\section*{Appendix: the forced harmonic oscillator}

To illustrate the surjectivity assumption in Section \ref{sec:criticalpoints}, consider the special case of a linear forced harmonic oscillator with potential $V(q)=(1/2)kq^2$ and dipole moment $D(q)=aq$.  Then, from Eqs. (\ref{eq:A}) and (\ref{eq:B}), we see that
\begin{equation*} A(t)=A=\left[\begin{matrix} 0 & \frac{1}{m} \\ -k & 0\\
\end{matrix}\right],\quad B(t)=B=\left[\begin{matrix}0\\ -a\\\end{matrix}\right],
\end{equation*}
which are both independent of time $t$. As a result, the matrix $M(t)$ satisfying Eq. (\ref{eq:Mdym}) can be simply expressed as $M(t)=\exp A t$, leading to the expression
\begin{equation*} \frac{\delta {\bf z}(T)}{\delta \epsilon(t)}=\left[\exp A(T-t)\right] B.\end{equation*}
We may explicitly determine $M(t)$ by noting that
\begin{equation*}
A = \begin{bmatrix} \frac{-i}{2\sqrt{km}} & -\frac{1}{2} \\ \frac{1}{2} & \frac{i\sqrt{mk}}{2}\end{bmatrix} \begin{bmatrix}i\sqrt{\frac{k}{m}} & 0 \\ 0 & -i\sqrt{\frac{k}{m}}\end{bmatrix}\begin{bmatrix}i\sqrt{km} & 1 \\ -1 & \frac{-i}{\sqrt{km}}\end{bmatrix}.
\end{equation*}
{\color{black} For ease of notation, we define the harmonic oscillator frequency as $\omega = \sqrt{\left(k/m\right)}$.  Then exponentiating  $A(T-t)$ and utilizing $B(t)$ above yields
\begin{align*}
\frac{\delta {\bf z}(T)}{\delta\epsilon(t)} = &\frac{-a}{2}\begin{bmatrix} \frac{-i}{\sqrt{km}}\big(\exp\left[i\omega(T-t)\right] - \exp\left[i\omega(t-T)\right]\big) \\ \exp\left[i\omega(T-t)\right] + \exp\left[i\omega(t-T)\right]\end{bmatrix} \\ = &-a\begin{bmatrix} \frac{1}{\sqrt{km}}\sin\left[\omega(T-t)\right] \\ \cos\left[\omega(T-t)\right]\end{bmatrix}
\end{align*}}
The two rows of $\delta{\bf z}(T)/\delta\epsilon(t)$ are linear independent functions of time, thereby showing that the surjectivity assumption holds for this system.

We further consider the special case of an anti-harmonic oscillator with potential $V(q)=-(1/2)kq^2$ ($k > 0$) and dipole moment $D(q)=aq$.  This system exhibits unstable behavior with ${\bf z}$ growing exponentially with time; nevertheless, it is a model for consideration of surjectivity analysis.  As with molecular systems where dissociation occurs, the dynamics would only be followed for a finite time $T$ {\color{black} out to} when the particles are separated beyond {\color{black} an ability to impose} further control.

Following the forced harmonic oscillator above, we may derive
{\color{black} \begin{equation*}
\frac{\delta {\bf z}(T)}{\delta\epsilon(t)} = -a\begin{bmatrix} \frac{1}{\sqrt{km}}\sinh\left[\omega(T-t)\right] \\ \cosh\left[\omega(T-t)\right]\end{bmatrix}.
\end{equation*}}
The two rows of $\delta{\bf z}(T)/\delta\epsilon(t)$ are linear independent functions of time, thereby showing that the surjectivity assumption holds for this extreme system.  

\end{document}